\documentclass{article}

\usepackage{PRIMEarxiv}

\usepackage[utf8]{inputenc} 
\usepackage[T1]{fontenc}    
\usepackage{hyperref}       
\usepackage{url}            
\usepackage{booktabs}       
\usepackage{amsfonts}       
\usepackage{nicefrac}       
\usepackage{microtype}      
\usepackage{lipsum}
\usepackage{fancyhdr}       
\usepackage{graphicx}       
\usepackage{tikz}
\usepackage{appendix}
\usepackage{subcaption}
\usepackage{amsmath}
\usepackage{float}

\title{Fusing Feature Engineering and Deep Learning: A Case Study for Malware Classification
}

\author{
  Daniel Gibert, Quan Le \\
  CeADAR \\
  University College Dublin \\
  Dublin, Ireland\\
  \texttt{\{daniel.gibert, quan.le\}@ucd.ie} \\
   \And
  Carles Mateu, Jordi Planes \\
  Polytechnic School \\
  University of Lleida \\
  Lleida, Spain\\
  \texttt{\{carles.mateu, jordi.planes\}@udl.cat} \\
}

\begin{document}
\maketitle

\begin{abstract}
Machine learning has become an appealing signature-less approach to detect and classify malware because of its ability to generalize to never-before-seen samples and to handle large volumes of data.
While traditional feature-based approaches rely on the manual design of hand-crafted features based on experts' knowledge of the domain, deep learning approaches replace the manual feature engineering process by an underlying system, typically consisting of a neural network with multiple layers, that perform both feature learning and classification altogether. However, the combination of both approaches could substantially enhance detection systems. In this paper we present 
an hybrid approach to address the task of malware classification by fusing multiple types of features defined by experts and features learned through deep learning from raw data. In particular, our approach 
relies on deep learning to extract N-gram like features from the assembly language instructions and the bytes of malware, and texture patterns and shapelet-based features from malware's grayscale image representation and structural entropy, respectively. These deep features are later passed as input to a gradient boosting model that combines the deep features and the hand-crafted features using an early-fusion mechanism. The suitability of 
our approach has been evaluated on the Microsoft Malware Classification Challenge benchmark and results show that the proposed solution achieves state-of-the-art performance and outperforms gradient boosting and deep learning methods in the literature.
\end{abstract}


\section{Introduction}
The fight against malware has never stopped since the dawn of computing. This fight has turned out to be a never-ending and cyclical arms race: as security analysts and researchers improve their defences, malware developers continue to innovate, find new infection vectors and enhance their obfuscation techniques. During the last decade, malware threats have been expanding vertically (i.e. numbers and volumes) and horizontally (i.e. types and functionality) due to the opportunities provided by technological advances~\footnote{\url{https://www.microsoft.com/en-us/security/business/security-intelligence-report}}. According to AV-Test~\footnote{\url{https://www.av-test.org/en/statistics/malware/}}, the total number of new malware detections worldwide amounted to 677.66 million programs, and it is projected to surprass 700 million within 2020.

Traditionally, anti-malware engines relied on signature-based and heuristic-based methods to detect and block malware before they performed any damage. On the one hand, signature-based methods identify malware by comparing its code with the code of known malware that have already encountered, analyzed and recorded in a database. 
On the other hand, heuristic-based methods examine the code and behavior of malware to look for certain malicious behaviors and suspicious properties. Most anti-malware engines that employ heuristic analysis run the program commands within a specialized virtual machine, to isolate the suspicious program from the real machine. Although effective, this type of analysis is very time consuming because it involves setting up a safe environment every time a suspicious file has to be analyzed.

As a result, the diversity, sophistication and availability of malicious software make the task of securing computer networks and systems very challenging, and force security analysts and researchers to constantly improve their cyberdefenses to keep pace with the attacks. During the last decade, due to the massive growth of malware streams, organizations faced the daunting challenge of dealing with thousands of attacks a day while also experiencing a shortage of cybersecurity skills and talent~\cite{csis2010human}. In consequence, new methods started to be adopted to complement traditional detection approaches and keep pace with new attacks and variants. This scenario presented a unique opportunity for machine learning, as an alternative to signature-based approaches, to impact and revamp the cybersecurity landscape, because of its ability to generalize to never-before-seen malware and to handle large volumes of data.

Traditional machine learning approaches for malware detection and classification rely on the manual extraction of hand-crafted features defined by experts in the domain~\cite{10.1145/2857705.2857713,7847046,7856826,Nataraj:2011:MIV:2016904.2016908,DBLP:conf/maics/MaysDB17}. However, these solutions depend almost entirely on the ability of the experts to extract characterizing features that accurately represent malware, and, the computational and memory requirements needed to extract some of these discriminant features, such as N-grams, limit their applicability in the real-world~\cite{Raff2016AnIO}. Lately, various approaches have been presented to automatically learn to extract N-gram like features from malware without having to exhaustively enumerate all N-grams during training using deep learning~\cite{21674592,GIBERT2021102159}, and approaches to automatically extract features from malware's structural entropy representation~\cite{AAAI1816133}, gray-scale image representation~\cite{Gibert2018,Khan2018}, from the binary code or any compressed representation of it~\cite{DBLP:conf/icann/GibertMP18,7966342}. For a complete review of feature-based and deep learning approaches we refer the readers to~\cite{GIBERT2020102526} and the references therein.

The task of malware detection and classification includes multiple types of features and thus, by only taking as input the raw bytes or opcodes sequence a great deal of useful information for classification is ignored such as the characteristics of the Portable Executable (PE) Headers, the import address table (IAT), etc. As a result, deep learning approaches tend to perform poorly in comparison to multimodal approaches in the literature~\cite{10.1145/2857705.2857713,7847046}. 

This paper presents an hybrid approach for malware classification that addresses the aforementioned limitation of deep learning approaches by combining deep and hand-crafted features using simple, but yet effective, early fusion mechanism to train gradient boosting trees~\cite{friedman2000greedy}
. As far as we know, this is the first approach to ever try to combine hand-crafted features defined by experts with features automatically learned through deep learning for the task of malware detection and classification. The main idea behind is to fuse both the deep and hand-crafted features into a single feature vector that is later used to train a single model to learn the correlation and interactions between each type of features. Our approach extracts well-known features such as API function calls, section characteristics, the frequency of usage of the registers, entropy features, and so on, and also extracts N-gram like features from the binary content and the assembly language source code of malware~\cite{21674592,GIBERT2021102159}, plus deep features from malware's structural entropy~\cite{AAAI1816133} and gray-scale image representation~\cite{Gibert2018}. This approach has been extensively assessed on the dataset provided by Microsoft for the Big Data Innovation Gathering Challenge of 2015~\cite{DBLP:journals/corr/abs-1802-10135}, which has become the de facto standard benchmark to evaluate machine learning models for malware classification. Results show a 99.81\% 10-fold cross validation accuracy on the training set and a 0.0040 logarithmic loss on the test set, outperforming any feature-based and deep learning-based approach in the literature. 

The rest of the paper is organized as follows. Section~\ref{sec:background} describes the methods employed to detect and classify malicious software.  Section~\ref{sec:related_work} introduces the related research to address the problem of malware detection and classification. Section~\ref{sec:methodology} provides a detailed description of our system, and the types of features extracted. Lastly, Section~\ref{sec:evaluation} describes the experimentation and Section~\ref{sec:conclusions} summarizes our research and presents some remarks on the ongoing research trends.

\section{Background}
\label{sec:background}
Next, it is introduced the necessary background required for the reader to understand malware forensics and the recently rise of machine learning to complement traditional detection methods.

\subsection{The Task of Malware Detection and Classification}
\label{sec:mlw_detection_and_classification_tasks}

Malicious software, also known as malware, is any kind of software that has been specifically designed to disrupt, harm or exploit any computer system or network. Typically, malware with similar characteristics and common behavior are grouped into families, whereas a malware family usually encompasses samples of malware that have been generated from the same code base. As malware keeps evolving, new variants or strains of a family might arise showing similar traits in their variations, just as these families have. These variations usually differ in key areas, such as those dealing with payload and infection. Therefore, the task of malware detection refers to the task of detecting whether a given piece of software is malicious or not while the task of malware classification refers to the task of distinguishing and classifying malware into families.  
Although the task of detecting malware is crucial to stop and prevent an attack before it causes damage, the task of malware classification helps to better understand how the malicious software has infected the system, its threat level and potential damage, and what can be done to defend against it and recover from the attack. 

\subsection{Traditional Detection Techniques}
\label{sec:traditional_detection_techniques}
Traditionally, to detect malware, anti-malware engines relied on signature and heuristic detection. On the one hand, signature detection involves comparing a program's code with the code of known malware that has already been encountered, analyzed and stored in a database, in order to find footprints matching those of known malware. Typically, if a program's code match one or more of those footprints, it is labelled as malicious and it will be either deleted or put into quarantine. For decades, this method of identifying malware has been the primary technique employed by anti-malware engines because of its simplicity. However, this detection technique solely works against known malware, limiting its protection against new forms of malware. In addition, modern malware can alter its signature to avoid detection by employing code obfuscation, dynamic code loading, encryption, packing, etc.
	
To counter the limitations of signature detection, anti-malware engines adopted heuristic detection. Oppositely to signature detection, which looks to match signatures or footprints found in known malicious files, heuristic detection uses rules and/or algorithms to look for pieces of code which may indicate malicious intent. For instance, a common technique is to look for specific commands or instructions that are not typically found in a benign application such as the payload of some trojan, the replication mechanism of virus, etc. Traditional heuristic anti-malware engines use some kind of rule-based or weight-based system to determine the maliciousness or the threat that a program poses to the computer system or network. If these rules exceed a predetermined threshold, then an alarm is triggered and a precautionary action is taken. However, these rules and heuristics have to be previously defined by experts after analyzing the behavior of malware, which is a complex and time consuming process, even for security experts.

A few decades ago, the number of threats attributed to malicious software was relatively low and simple hand-crafted rules were sufficient to detect the ongoing threats. However, during the last decade, malware has exploded in terms of diversity, sophistication, and availability, and thus, the aforementioned detection and analysis techniques have been unable to keep pace with the new attacks and variants. In addition, the shortage of experienced security researchers and analysts~\cite{csis2010human} plus the complexity of forensic investigations have contributed to strengthening the use of machine learning as an 
appealing signature-less alternative to detect malware because of (1) its ability to generalize to never-before-seen malware and (2) to handle large volumes of data.

\section{Related Work}
\label{sec:related_work}
In this section, the related studies performed in the field of malware detection and classification that are powered by machine learning are presented. For a complete review of the features and deep learning architectures defined by experts to build machine learning detection systems we refer the readers to Ucci et al.~\cite{UCCI2019123}, Gibert et al.~\cite{GIBERT2020102526}, and the references therein.
	
Traditional machine learning approaches for malware detection and classification rely on the manual extraction of features defined by security experts~\cite{10.1145/2857705.2857713,7847046,7856826,Nataraj:2011:MIV:2016904.2016908,DBLP:conf/maics/MaysDB17}. Feature extraction is one key step to build a malware detection and classification system. It transforms raw data, i.e. binary executables, into numerical features that provide an abstract representation of the original data. 
Common features are byte and opcode N-grams, API function calls, the frequency of use of registers, characteristics extracted from the header and sections of executables, etcetera.
However, these solutions depend almost entirely on the ability of the security experts to extract a set of descriptive features that accurately represent the malware characteristics. In addition, the computational and memory requirements needed to extract some of these features, such as N-grams, far exceeds the capabilities of most existing systems and limit their applicability in the real-world~\cite{Raff2016AnIO}.
	
Lately, various approaches have been presented to automatically learn to extract N-gram like features from malware without having to exhaustively enumerate all N-grams during training through deep learning. For instance,
Gibert et al.~\cite{21674592,GIBERT2021102159} proposed a shallow convolutional neural network architecture that intrinsically learn to extract N-gram like features from both the malware's binary content and its assembly language source code, respectively.
This is achieved by a convolutional layer with filters of various sizes followed by a global max-pooling layer, which allow the model to retrieve the features regardless of their location in the assembly language instructions and byte sequences. Recently, other approaches~\cite{DBLP:conf/aaai/RaffBSBCN18,krcal2018deep} have been proposed to detect malware based on their binary content. 
Raff et al.~\cite{DBLP:conf/aaai/RaffBSBCN18} proposed a shallow convolutional neural network architecture with a gated convolutional layer to capture the high location invariance in malware's binary content while Krčál et al.~\cite{krcal2018deep} presented the deepest architecture to date, consisting of an embedding layer followed by two convolutions, one max-pooling layer, two more convolutions, and a global average pooling layer and four fully connected layers.

However, dealing with raw byte sequences directly is very challenging as it involves the classification of sequences of extreme lengths. For instance, the size of the executables may range from some KBs to hundreds of MBs, e.g. 100 MB correspond to a sequence of 100,000,000 bytes. To deal with these sequences of extreme length, various approaches~\cite{7966342,AAAI1816133,Gibert2018,DBLP:conf/icann/GibertMP18,Khan2018,10.1007/978-3-030-36711-4_14,GAO2020102661,XIAO2021102420,SUDHAKAR2021334} proposed to compress the information in the malware's binary content.
For instance, Gibert et al.~\cite{AAAI1816133} presented an approach to classify malware represented as a stream of entropy values. In their work, the binary content is divided into chunks of code of fixed size and afterwards, the information at each chunk is compressed by calculating its entropy value. 
On the other hand, executables could be represented as grayscale images~\cite{Gibert2018,Khan2018,XIAO2021102420,SUDHAKAR2021334}. To represent a malware sample as a grayscale image, every byte has to be interpreted as one pixel in an image, where values are in the [0,255] (0:black, 255:white). Afterwards, the resulting 1-D array has to be reorganized as a 2-D array.

Nevertheless, the task of malware detection and classification is regarded as a multimodal task, as it includes multiple types of features and thus, by only taking as input the assembly language instructions and the raw bytes sequence or a compressed representation of it, a lot of useful information for characterizing malware is ignored. As a result, these unimodal deep learning approaches that only take as input a single source of information tend to perform poorly in comparison to multimodal approaches~\cite{10.1145/2857705.2857713,7847046,electronics9050721,9058025} that extract different types of features from various modalities of information. This is because these types of features or modalities provide complementary information to each other, and reflect patterns not visible when working with individual modalities on their own. In fact, these approaches that extract multiple types of features from malware remained unbeaten in terms of classification performance and have been the way to go for detecting and classifying malware.

To solve the aforementioned issues of deep learning approaches, we present a multimodal system that combines the benefits of feature engineering and deep learning to achieve state-of-the-art results in the task of malware detection and classification.

\section{System Overview}
\label{sec:methodology}
\begin{figure*}[ht]
	\centering
	\includegraphics[width=\textwidth]{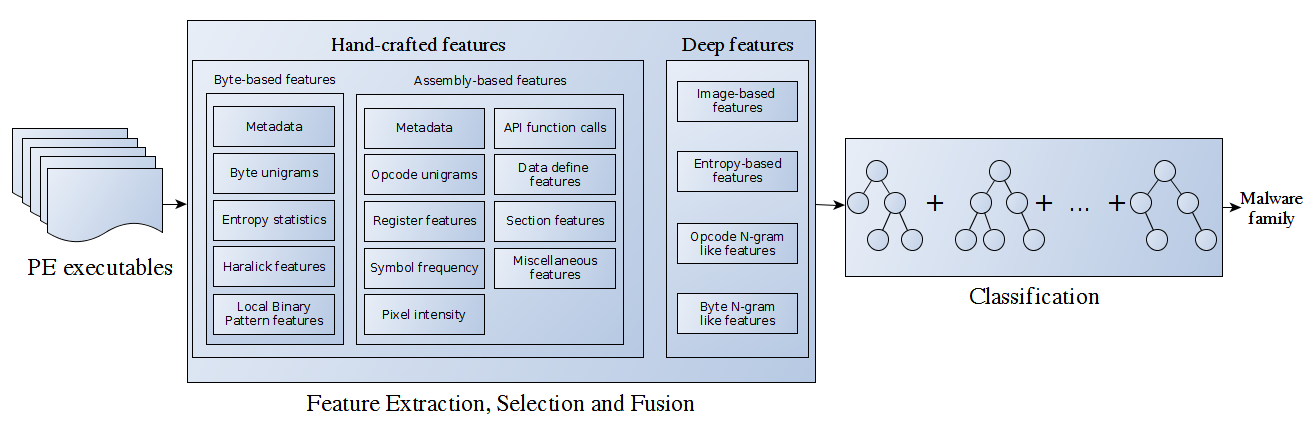}
	\caption{Overview of the proposed malware classification system.}
	\label{fig:system_overview}
\end{figure*}

The proposed classification system extracts hand-engineered and deep features to address the task of malware classification by combining the aforementioned features using a simple, but yet effective, early fusion mechanism in order to train gradient boosting trees~\cite{friedman2000greedy}
. Following, the components of the proposed system are described in detail. First, the different types of features extracted are explained. Then, the deep learning architectures implemented to automatically learn and extract deep features from raw data are introduced. Afterwards, it is described the proposed fusion mechanism to combine the aforementioned features. Lastly, it is introduced the machine learning algorithm trained to categorize malware into families. See Figure~\ref{fig:system_overview} for an overview of the system. 
\subsection{Hand-engineered Features}
This system has been implemented to tackle the problem of Windows malware classification and thus, the features have been specifically designed to be extracted from executables in the Portable Executable (PE) file format, which is the format for executables, DLLs, FON Font files and others in the Windows operating system. 

\begin{figure}[ht]
	\centering
	\includegraphics[width=0.5\columnwidth]{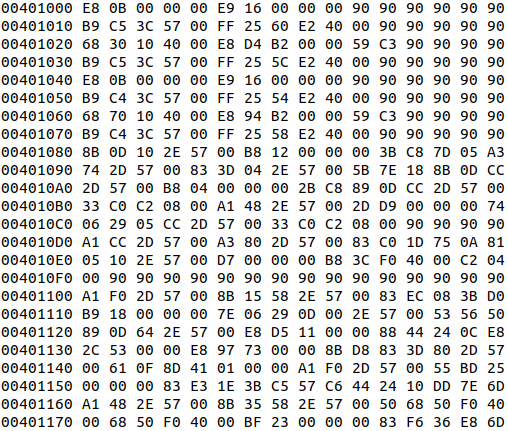}
	\caption{Hexadecimal view of malware.}
	\label{fig:hex_view}
\end{figure}

\begin{figure}[ht]
	\centering
	\includegraphics[width=0.5\columnwidth]{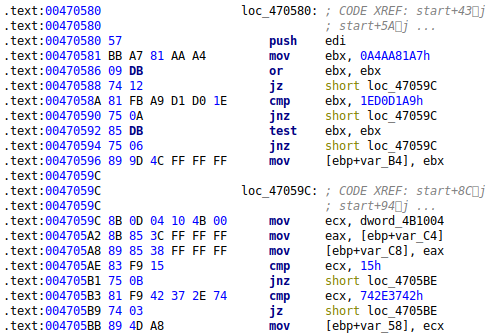}
	\caption{Assembly view of malware.}
	\label{fig:assembly_view}
\end{figure}

There are two common representations of a Windows malware executable: (1) the hexadecimal representation of malware's binary content (or hex view) and (2) the assembly language source code of malware (or assembly view). On the one hand, the hex view of malware represents the machine code as a sequence of hexadecimal values. Cf. Figure~\ref{fig:hex_view}.
The first value indicates the starting address of the machine codes in the memory, and each hexadecimal value (byte) carry meaningful information of the Portable Executable file such as instruction codes and data. On the other hand, the assembly language source code contains the symbolic machine code of the executable, i.e. the machine code instructions, as well as function calls, memory allocation and variable information. Cf. Figure~\ref{fig:assembly_view}. To obtain the source code, malware binaries have to be first disassembled. Typically, it is used the Interactive Disassembler (IDA) tool~\footnote{\url{https://www.hex-rays.com/products/ida/}}, Radare2~\footnote{\url{https://rada.re/n/}} or Ghydra~\footnote{\url{https://www.nsa.gov/resources/everyone/ghidra/}}, among others.
To build an accurate classifier, we propose to extract hand-crafted features from both representations, the hex view and the assembly view of the executables, to exploit the complementary information provided by these two representations. More specifically, we decided to limit our proposed approach to only common and well-known features~\cite{10.1145/2857705.2857713,7847046} in order to evaluate whether or not the classification performance of the system improves after the addition of deep features extracted by the deep learning models. Next, the hand-crafted features extracted from the hexadecimal source code of malware's binary content and the assembly language source code are presented.
\subsubsection{Hexadecimal-based features}
\label{sec:manual_hex_features}
Hexadecimal-based features refer to those features extracted from the hexadecimal representation of malware's binary content. Following you will find a brief description of all the hexadecimal-based feature types used by our system.
\begin{description}
	\item[Metadata information (\emph{BYTE\_MD}).] Two features compose the metadata information extracted from the hexadecimal view of malware, (1) the size of the file and (2) the address of the first byte sequence.
	\item[Byte unigram features (\emph{BYTE\_1G}).] Remember that in a byte sequence, each element can take one out of 256 different values, i.e. ranging from 0 to 255 (the byte range). In addition, some elements are represented by the special symbol ??, indicating that the corresponding byte has no mapping in the executable file (the contents of those addresses are uninitialized within the file). However, better results can be achieved by only taking as input the 256 byte values~\cite{10.1145/2857705.2857713}. For this reason, we decided to leave the ?? out the experiments.
	\item[Entropy-based features (\emph{BYTE\_ENT}).] Entropy has long been used in the security industry to detect the presence of encrypted and compressed code as these tend to have higher entropy than native code~\cite{4140989}. To sum up, the entropy of a byte sequence refers to the amount of disorder of the distribution of bytes, whose value range from 0 (order) to 8 (randomness). Generally speaking, if occurrences of all byte values are the same, the entropy will be larger, but if certain byte values occur with higher probability, the entropy will be smaller. Nevertheless, the use of simple entropy statistics may not be enough to detect malware as authors usually try to conceal the encrypted and compressed code in a way that they bypass high entropy filters. For this reason, the bytes sequence is typically represented as a stream of entropy values (aka structural entropy). See Figure~\ref{fig:structural_entropy_sample}. To calculate the structural entropy of an executable, the bytes sequence is split into non-overlapping chunks of fixed size, e.g. 256, and for each chunk of code, its entropy value is computed as follows:
	\begin{equation}
	H(X) = - \sum_{i=1}^{n} p(i)\cdot log_{b}p(i)
	\end{equation}
	where $H(X)$ is the measured entropy value of a given chunk of code $X$ with values ${x_{1},\ldots,x_{j}}$, $j$ is the number of values in X, $p(i)$ refers to the probability of appearances of the byte value $i$ in $X$ and $n$ is equal to $255$, i.e. byte code values are in the range of [0, 255].
	
	\begin{figure}[ht]
		\centering
		\includegraphics[width=0.6\columnwidth]{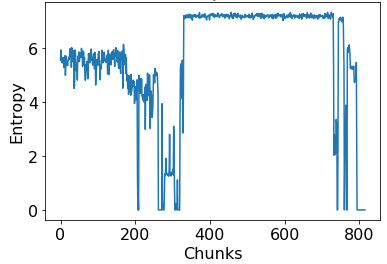}
		\caption{Structural entropy of a malware sample.}
		\label{fig:structural_entropy_sample}
	\end{figure}
	
	Consequently, we extracted various statistical features from the structural entropy of executables such as the mean, variance, median, maximum, minimum entropy values, and the percentiles.
	\item[IMG-based features (\emph{BYTE\_HARALICK}, \emph{BYTE\_LBP}).] Nataraj et al.~\cite{10.1145/2016904.2016908} proposed a method for visualizing and classifying malware using image processing techniques. In their work, malware binaries are visualized as grayscale images, with every byte reinterpreted as one pixel in the image. Then, the resulting array is reorganized as a 2-D array and visualized as a grayscale image, where values are in the range [0, 255], 0 for black and 255 for white. The rational behind this representation is that images of malicious software from a given family are similar between them but distinct from those belonging to a different family and thus, this visual similarity can be exploited to classify malware. Cf. Figure~\ref{fig:grayscale_img_malware_samples}.
	
	\begin{figure}[ht]
		\centering
		\includegraphics[width=0.9\columnwidth]{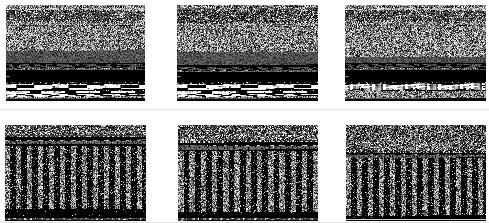}
		\caption{Grayscale image representation of samples belonging to the Obfuscator.ACY and Gatak families, respectively. Notice that the images of malicious software belonging to the same family are similar between them while distinct from those images of malware belonging to the other family.}
		\label{fig:grayscale_img_malware_samples}
	\end{figure}
	
	From these visual representation, we extracted Haralick~\cite{4309314} and Local Binary Pattern~\cite{576366} features. On the one hand, Haralick features are calculated by constructing a co-occurence matrix and computing the Haralick equations, e.g. angular second-moment, contrast, correlation, etc. On the other hand, the local binary pattern features of a given pixel are computed as follows. First, a 8 bit binary array is initialized with all values equal to 0. Then, the given pixel is compared with its neighboring pixels in clockwise direction. If the value of the neighboring pixel is greater or equal to 1 is assigned to its corresponding position.  This results in a 8 bit binary array with zeros and ones. Following, the 8-bit binary pattern is converted to a decimal number and it is stored in the corresponding pixel location in the LBP mask. Finally, this process is applied to all pixels in the image and, once the LBP values of all pixels have been calculated, the mask is normalized, resulting in 256 features.
\end{description}

\subsubsection{Assembly-based features}
Assembly-based features refer to those features extracted from the assembly language source code of malware. Following you will find a brief description of the the assembly-based feature types used by our system.

\begin{description}
	\item[Metadata information (\emph{ASM\_MD}).] Two features compose the metadata information extracted from the assembly language source code of malware, (1) the size of the file and (2) the number of lines in the file.
	\item[Operation codes unigram features (\emph{ASM\_OPC}).] An operation code or opcode is the portion of a machine language instruction that specifies the operation to be performed, e.g. \emph{ADD}, \emph{MUL}, \emph{SUB}, etc. Instead of calculating the frequency of all opcodes, we just selected a subset of 93 opcodes based on their commonness and frequency of use in malicious applications~\cite{danielbilarbib}. The full list of opcodes extracted from each assembly language source code file can be found in the project's repository~\footnote{\label{https://github.com/danielgibert/malware_classification_with_gradient_boosting_and_deep_features}}.
	\item[Data define features (\emph{ASM\_DD}).] Data define directives are used for reserving storage for variables. However, because of packing, some malware samples do not contain any API call and barely a few operation codes. More specifically, those samples of malware mostly contain db, dw and dd define directives, which are used for setting byte, word, and double words, respectively. Subsequently, the frequency of use of the aforementioned data define directives has high discriminative power for a number of malware families. The full list of data define features can be found in Table~\ref{tab:data_define_directive_features}.
	\begin{table*}[ht]
		\centering
		\caption{List of features in the~\emph{ASM\_DD} category~\cite{10.1145/2857705.2857713}.}
		\label{tab:data_define_directive_features}
		\resizebox{\textwidth}{!}{%
			\begin{tabular}{ll}
				\hline
				\multicolumn{1}{l|}{Feature Name}   & Description                                                                                                                    \\ \hline
				\multicolumn{1}{l|}{db\_por}        & Proportion of the db data define directive to the whole file                                                                          \\
				\multicolumn{1}{l|}{dd\_por}        & Proportion of the dd data define directive to the whole file                                                                          \\
				\multicolumn{1}{l|}{dw\_por}        & Proportion of the dw data define directive to the whole file                                                                          \\
				\multicolumn{1}{l|}{dc\_por}        & Proportion of the db, dd and dw data define directives to the whole file                                                              \\
				\multicolumn{1}{l|}{db0\_por}       & Proportion of the db data define directive with 0 parameters to the whole file                                                        \\
				\multicolumn{1}{l|}{dbN0\_por}      & Proportion of the db data define directive with more than 0 parameters to the whole file                                              \\
				\multicolumn{1}{l|}{db\_text}       & Proportion of the db data define directive in the .text section                                                                       \\
				\multicolumn{1}{l|}{db3\_rdata}   & Proportion of the db data define directive with one non-zero parameter to the .rdata section                                          \\
				\multicolumn{1}{l|}{db3\_data}      & Proportion of the db data define directive with one non-zero parameter to the .data section                                           \\
				\multicolumn{1}{l|}{db3\_idata}                          & Proportion of the db data define directive with one non-zero parameter to the .idata section                                          \\
				\multicolumn{1}{l|}{db3\_all}       & Proportion of the db data define directive with one non-zero parameter to the whole file                                              \\
				\multicolumn{1}{l|}{db3\_NdNt}      & Proportion of the db data define directive with one non-zero parameter in the unknown sections                                        \\
				\multicolumn{1}{l|}{db3\_zero\_all} & Proportion of the db data define directive with 0 parameter with respect to the number of db data define directives with non-zero parameters \\
				\multicolumn{1}{l|}{dd\_text}       & Proportion of the dd data define directive in the text section                                                                        \\
				\multicolumn{1}{l|}{dd\_rdata}      & Proportion of the dd data define directive to the .rdata section                                                                      \\
				\multicolumn{1}{l|}{dd4}            & Proportion of the dd data define directive with four parameters                                                                       \\
				\multicolumn{1}{l|}{dd5}            & Proportion of the dd data define directive with five parameters                                                                       \\
				\multicolumn{1}{l|}{dd6}            & Proportion of the dd data define directive with six parameters                                                                        \\
				\multicolumn{1}{l|}{dd4\_all}       & Proportion of the dd data define directive with four parameters to the whole file                                                     \\
				\multicolumn{1}{l|}{dd5\_all}       & Proportion of the dd data define directive with five parameters to the whole file                                                     \\
				\multicolumn{1}{l|}{dd6\_all}       & Proportion of the dd data define directive with six parameters to the whole file                                                      \\
				\multicolumn{1}{l|}{dd4\_NdNt}      & Proportion of the dd data define directive with four parameters in unknown sections                                                   \\
				\multicolumn{1}{l|}{dd5\_NdNt}      & Proportion of the dd data define directive with five parameters in unknown sections                                                   \\
				\multicolumn{1}{l|}{dd6\_NdNt}      & Proportion of the dd data define directive with six parameters in unknown sections   \\ \hline                                                
			\end{tabular}
		}
	\end{table*}
	
	\item[Register features (\emph{ASM\_REG}).] Registers are like variables built in the processor. Using registers instead of memory to store values makes the process faster and cleaner. The x86 architecture of processors contains eight General-Purpose Registers. 
	\begin{table}[ht]
		\centering
		\caption{16-bit naming convention of the eight General-Purpose Registers (GPR).}
		\label{tab:gpr}
			\begin{tabular}{l|l}
				\hline
				Register type                             & Description                                     \\ \hline
				Accumulator register (AX)        & Employed in arithmetic operations (e.g. INC, DEC, ADD, SUB, etc).   \\
				Counter register (CX)            & Employed in shift/rotate instructions and loops (e.g. JMP, JNZ, etc) \\
				Data register (DX)               & Employed in arithmetic and I/O operations (e.g. IN, INS, OUT, etc)           \\
				Base register (BX)               & Employed as a pointer to data                       \\
				Stack Pointer register (SP)      & Pointer to the top of the stack                 \\
				Stack Base Pointer register (BP) & Pointer to the base of the stack                \\
				Source Index register (SI)       & Pointer to the source in stream operations      \\
				Destination Index register (DI)  & Pointer to the destination in stream operations \\ \hline
			\end{tabular}%
	\end{table}
	All registers can be accessed in 16-bit and 32-bit modes. In 16-mode, the registers are abbreviated using the abbreviations listed in Table~\ref{tab:gpr}. In 32-bit mode, it is added the prefix 'E' to this two-letter abbreviation. For instance, 'ECX' it the counter register as a 32-bit value. Similarly, in the 64-bit version, the 'E' is replaced with an 'R'. Thus, the 64-bit version of 'ECX' is 'RCX'.
	In addition, the first four registers, AX, CX, DX and BX, can be addressed as two 8-bit halves~\footnote{\url{https://en.wikibooks.org/wiki/X86_Assembly/X86_Architecture\#x86_Architecture}}.
	Moreover, there are six Segment registers used to store the starting addresses of the code, the data and the stack segments, namely the Stack Segment, the Code Segment, the Data Segment, the Extra Segment, the F Segment and the G Segment. 
	
	\item[Symbols frequency (\emph{ASM\_SYM}).] The usage of the following set of symbols -, +, *, ], [, ?, @ are extracted as features because these characters are typically found in code that has been designed to evade detection by resorting to indirect calls and to dynamic library loading. 
	\item[Application Programming Interface (ASM\_API).] Application Programming Interface (API) functions and system calls are related to services provided by the operating systems, in our case Windows. These functions provide access to key operations such as network, security, system services, file management, and so on, and there are the only way for software to access system resources managed by the operating system. As a result, API function calls provide descriptive information with respect to the intent or behavior of a particular piece of software. For this reason, we measured the frequency of use of a subset of 794 API functions based on their frequency on a study of nearly 500 thousand malicious samples~\cite{api_subset}. The rational behind using only a subset of API functions is that the total number of APIs is extremely large and considering all of them would bring little to no meaningful information to the task of malware classification. The subset of API functions extracted from the assembly language source code file can be found in the project's repository
	\footnote{\ref{https://github.com/danielgibert/malware_classification_with_gradient_boosting_and_deep_features}}.
	\item[Pixel intensity features (\emph{ASM\_PIXEL}).]  Just as malware's binary content, the assembly language source code of malware can also be visualized as a grayscale image. Cf. Figure~\ref{fig:asm_images}. In fact, the Winner's of the Big Data Innovators Gathering Challenge of 2015~\footnote{\url{https://www.kaggle.com/c/malware-classification}}, and state-of-the-art approaches in the literature~\cite{7847046} have shown that the intensities of pixels in the assembly-based grayscale images work well when used in conjunction with other features. 
	\begin{figure}[ht]
		\centering
		\includegraphics[width=\columnwidth]{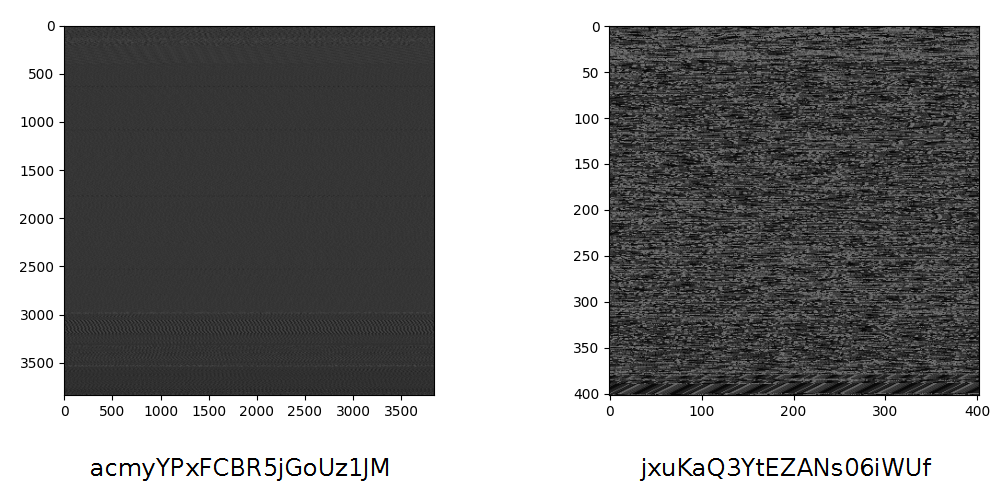}
		\caption{Grayscale image representation of malware's assembly language source code.}
		\label{fig:asm_images}
	\end{figure}
	
	\item[Section features (\emph{ASM\_SEC}).] A Portable Executable (PE) file consists of various predefined sections, e.g. .text, .data, .bss, .rdata, .edata, .idata, .rsrc, .tls, and .reloc. However, malware authors usually employ obfuscation techniques such as packing, which might modify the default sections and create new sections. For example, the UPX packer (Ultimate Packer for Executables), one of the most popular and well-known packer, typically creates two sections named UPX0 and UPX1. For this reason, we extracted different characteristics from the sections in the PE executables related to the proportion of code in those sections with respect to the whole file. The full list of section-based features is presented in Table~\ref{tab:section_features}.
	\begin{table}[ht]
		\centering
		\caption{List of features in the~\emph{ASM\_SEC} category~\cite{10.1145/2857705.2857713}.}
		\label{tab:section_features}
			\begin{tabular}{l|l}
				\hline
				Feature Name             & Description                                        \\ \hline
				.bss                     & Total number of lines in the .bss section          \\
				.data                    & Total number of lines in the .data section         \\
				.edata                   & Total number of lines in the .edata section        \\
				.idata                   & Total number of lines in the .idata section        \\
				.rdata                   & Total number of lines in the .rdata section        \\
				.rsrc                    & Total number of lines in the .rsrc section         \\
				.text                    & Total number of lines in the .text section         \\
				.tls                     & Total number of lines in the .tls section          \\
				.reloc                   & Total number of lines in the .reloc section        \\
				.bss\_por                & Proportion of .bss section to the whole file       \\
				.data\_por               & Proportion of the .data section to the whole file  \\
				.edata\_por              & Proportion of the .edata section to the whole file \\
				.idata\_por              & Proportion of the .idata section to the whole file \\
				.rdata\_por              & Proportion of the .rdata section to the whole file \\
				.rsrc\_por               & Proportion of the .rsrc section to the whole file  \\
				.text\_por               & Proportion of the .text section to the whole file  \\
				.tls\_por                & Proportion of the .tls section to the whole file   \\
				.reloc\_por              & Proportion of the .reloc section to the whole file \\
				Num\_Section             & Total number of sections                           \\
				Unknown\_Sections        & Total number of unknown sections                   \\
				Known\_Sections\_lines   & Total number of lines in known sections            \\
				Unknown\_Sections\_lines & Total number of lines in unknown sections          \\
				Known\_Sections\_por     & Proportion of known sections to all sections       \\
				Unknown\_Sections\_por   & Proportion of the unknown sections to all sections \\ \hline
			\end{tabular}
	\end{table}
	\item[Miscellaneous features (\emph{ASM\_MISC}).] This feature category consists of the frequency of 95 manually chosen keywords from the assembly language source code~\cite{10.1145/2857705.2857713}, mostly consisting of strings and dlls. 
	The full list of miscellaneous features is listed in the project's repository
	\footnote{\ref{https://github.com/danielgibert/malware_classification_with_gradient_boosting_and_deep_features}}.
\end{description}

\subsection{Deep Learning Architectures}
\label{sec:deep_architectures}
Our system also automatically extracts features from raw data through deep learning. More specifically, it relies on the extraction of N-gram like features from both the hexadecimal representation of malware's binary content (byte N-grams) and its assembly language source code (opcode N-grams), shapelet-based features from malware's structural entropy, and texture patterns from malware's binary content represented as grayscale images.
\subsubsection{Texture-based Features from the Grayscale Image Representation of Malware}
Instead of using well-know feature extractors, e.g. Haralick~\cite{4309314}, Local Binary Patterns~\cite{576366}, etc, we implemented a convolutional neural network to automatically learn discriminative features (\emph{BYTE\_IMG\_CNN}) from the grayscale image representation of malware~\cite{10.1145/2016904.2016908} given its superior performance for the task of malware classification~\cite{Gibert2018}. In their work, Gibert et al.~\cite{Gibert2018} compared the performance of CNN classifiers against various state-of-the-art feature extractors, including Haralick, Local Binary Pattern, PCA, and GIST, for the task of malware classification and the results show a 2.31\%, 1.88\%, 1.99\%, and a 1.34\% increase with respect to the classification performance of the classifiers trained with the LBP, Haralick, PCA and GIST features, respectively.

\begin{figure}[ht]
	\centering
	\tikzstyle{state}=[circle,
	thick,
	minimum size=1.2cm,
	draw=black,
	fill=white]
	\tikzstyle{input_output} = [rectangle, rounded corners, minimum width=3cm, minimum height=1cm,text centered, draw=black, fill=red!30]
	\tikzstyle{relu_act} = [rectangle, minimum width=1cm, minimum height=1cm,text centered, draw=black, fill=green!20]
	\tikzstyle{selu_act} = [rectangle, minimum width=1cm, minimum height=1cm,text centered, draw=black, fill=cyan!10]
	\tikzstyle{softmax} = [rectangle, minimum width=3cm, minimum height=1cm,text centered, draw=black, fill=red!30]
	\tikzstyle{pooling} = [rectangle, minimum width=5cm, minimum height=1cm,text centered, draw=black, fill=cyan!10]
	\tikzstyle{relu} = [rectangle, minimum width=5cm, minimum height=1cm,text centered, draw=black, fill=green!20]
	\tikzstyle{fully} = [rectangle, minimum width=5cm, minimum height=1cm,text centered, draw=black, fill=yellow!10]
	\tikzstyle{nonactivation} = [rectangle, minimum width=3cm, minimum height=1cm,text centered, draw=black, fill=white]
	
	\tikzstyle{embedding} = [rectangle, minimum width=5cm, minimum height=1cm,text centered, draw=black, fill=yellow!30]
	\tikzstyle{normalization} = [rectangle, minimum width=5cm, minimum height=1cm,text centered, draw=black, fill=gray!20]
	
	\begin{tikzpicture}[node distance=1.5cm]
	\node (grayscale_img) [input_output,label={left:$255 \times 255$},xshift=1.5] {Grayscale image};
	\node (conv_1) [relu, below of=grayscale_img, label={left:$50$}] {Conv 2D size $5 \times 5$ (strides 1)};
	\node (pool_1) [pooling, below of=conv_1, yshift=0.5cm] {Max-pooling 2D size $2 \times 2$};
	\node (normalization_1) [normalization, below of=pool_1, yshift=0.5cm] {Layer normalization};
	
	\node (conv_2) [relu, below of=normalization_1, label={left:$50$}] {Conv 2D size $3 \times 3$ (strides 1)};
	\node (pool_2) [pooling, below of=conv_2, yshift=0.5cm] {Max-pooling 2D size $2 \times 2$};
	\node (normalization_2) [normalization, below of=pool_2, yshift=0.5cm] {Layer normalization};
	
	\node (conv_3) [relu, below of=normalization_2, label={left:$70$}] {Conv 2D size $3 \times 3$ (strides 1)};
	\node (pool_3) [pooling, below of=conv_3, yshift=0.5cm] {Max-pooling 2D size $2 \times 2$};
	\node (normalization_3) [normalization, below of=pool_3, yshift=0.5cm] {Layer normalization};
	
	\node (fully_1) [fully, below of=normalization_3, label={left:$256$}] {Fully-connected layer};
	\node (output) [input_output, below of=fully_1, label={left:$9$}] {Output layer};
	
	\draw[->] (grayscale_img) -- (conv_1);
	\draw[->] (normalization_1) -- (conv_2);
	\draw[->] (normalization_2) -- (conv_3);
	\draw[->] (normalization_3) -- (fully_1);
	\draw[->] (fully_1) -- (output);
	\end{tikzpicture}
	\caption{Convolutional neural network architecture for classifying malware's binary content represented as a grayscale image.}
	\label{fig:cnn_grayscale_img}
\end{figure}
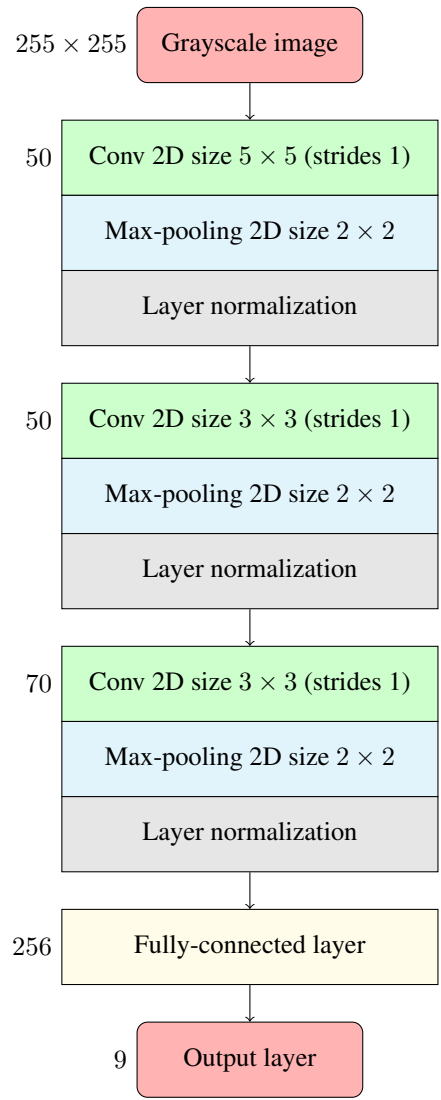

The architecture of our choice is the one used in the work of Gibert et al.~\cite{Gibert2018}. In particular, this architecture consists of three convolutional blocks composed by a convolutional layer, max-pooling layer and a normalization layer, followed by two fully-connected layers and a softmax layer. See Figure~\ref{fig:cnn_grayscale_img}. Notice that the resulting grayscale images have been resized to size equals $255 \times 255$ using the Lanzcos filter as in their work. For more details about the architecture we refer the reader to the original article~\cite{Gibert2018}.

\subsubsection{Shapelet-based Features from Malware's Structural Entropy Representation}
\begin{figure}[ht]
	\centering
	\tikzstyle{state}=[circle,
	thick,
	minimum size=1.2cm,
	draw=black,
	fill=white]
	\tikzstyle{input_output} = [rectangle, rounded corners, minimum width=3cm, minimum height=1cm,text centered, draw=black, fill=red!30]
	\tikzstyle{relu_act} = [rectangle, minimum width=1cm, minimum height=1cm,text centered, draw=black, fill=green!20]
	\tikzstyle{selu_act} = [rectangle, minimum width=1cm, minimum height=1cm,text centered, draw=black, fill=cyan!10]
	\tikzstyle{softmax} = [rectangle, minimum width=3cm, minimum height=1cm,text centered, draw=black, fill=red!30]
	\tikzstyle{pooling} = [rectangle, minimum width=5cm, minimum height=1cm,text centered, draw=black, fill=cyan!10]
	\tikzstyle{relu} = [rectangle, minimum width=5cm, minimum height=1cm,text centered, draw=black, fill=green!20]
	\tikzstyle{fully} = [rectangle, minimum width=5cm, minimum height=1cm,text centered, draw=black, fill=yellow!10]
	\tikzstyle{nonactivation} = [rectangle, minimum width=3cm, minimum height=1cm,text centered, draw=black, fill=white]
	
	\tikzstyle{embedding} = [rectangle, minimum width=5cm, minimum height=1cm,text centered, draw=black, fill=yellow!30]
	\tikzstyle{normalization} = [rectangle, minimum width=5cm, minimum height=1cm,text centered, draw=black, fill=gray!20]
	\begin{tikzpicture}[node distance=1.5cm]
	\node (structural_entropy) [input_output, label={left:$N$}] {Structural Entropy};
	
	\node (conv_1) [relu, below of=structural_entropy, label={left:$50$}] {Conv 1D size 3 (Stride 1)};
	
	\node (maxpooling_1) [pooling, below of=conv_1, yshift=0.5cm] {Max-pooling 1D size 2};
	
	\node (conv_2) [relu, below of=maxpooling_1, label={left:$70$}] {Conv 1D size 3 (stride 1)};
	\node (maxpooling_2) [pooling, below of=conv_2, yshift=0.5cm] {Max-pooling 1D size 2};
	
	\node (conv_3) [relu, below of=maxpooling_2, label={left:$70$}] {Conv 1D size 3 (stride 1)};
	\node (maxpooling_3) [pooling, below of=conv_3, yshift=0.5cm] {Max-pooling 1D size 2};

	\node (fully1) [fully, below of=maxpooling_3, label={left:$1000$}] {Fully-connected layer};
	\node (fully2) [fully, below of=fully1, label={left:$300$}, yshift=0.5cm] {Fully-connected layer};
	
	\node (softmax) [input_output, below of=fully2, label={left:$9$}] {Output layer};
	
	\draw[->] (structural_entropy) -- (conv_1);
	\draw[->] (maxpooling_1) -- (conv_2);
	\draw[->] (maxpooling_2) -- (conv_3);
	\draw[->] (maxpooling_3) -- (fully1);
	\draw[->] (fully2) -- (softmax);
	
	\end{tikzpicture}
	\caption{Convolutional neural network architecture for classifying malware's binary content represented as a stream of entropy values.}
	\label{fig:sructural_entopy_cnn}
\end{figure}
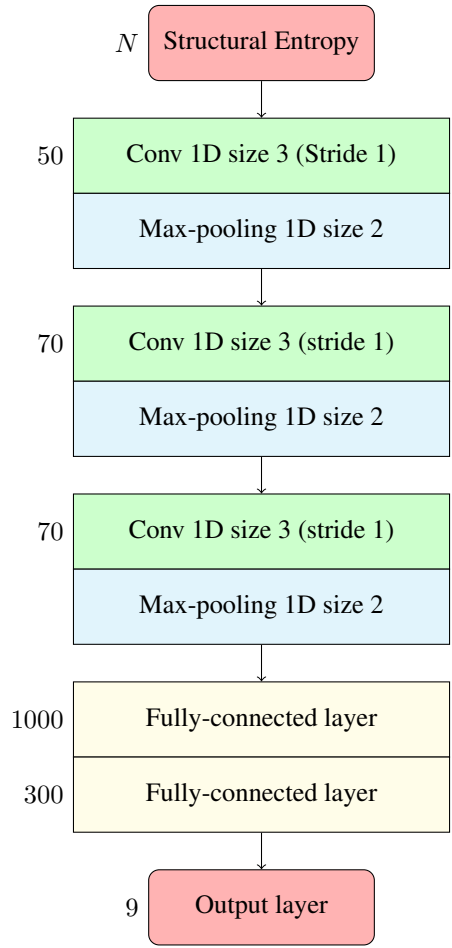

To complement the statistical hand-crafted features (\emph{BYTE\_ENT}) defined in Section~\ref{sec:manual_hex_features} we implemented a convolutional neural network architecture 
to learn features from the structural entropy representation of malware's binary content~\cite{AAAI1816133}. Cf. Figure~\ref{fig:sructural_entopy_cnn}. As it has been documented in Gibert et al.~\cite{AAAI1816133} and Baysa et al.~\cite{DBLP:journals/virology/BaysaLS13}, the structural entropy of malware belonging to the same family is very similar while distinct from the structural entropy representation of malware belonging to the other families. The rationale behind applying convolutional layers to the stream of entropy values is to learn hierarchical shapelet-like features from this kind of representation. A shapelet is a subsequence of a time series which is representative of a class (family). The idea behind is to distinguish the samples of malware belonging to different families by their local variations instead of their global structure However, brute-force shapelet-based approaches are computationally expensive~\cite{10.1145/2623330.2623613} and thus, we used convolutional layers to learn which are the optimal shapelets without exploring all possible candidates. More specifically, the convolutional layers can be seen as detection filters for specific subsequences present in the structural entropy of malware, going from low-level features in the first layers to increasingly complex features in the last layers. 

\subsubsection{Byte and Opcode N-Gram like Features}
One of the most common types of features to classify malware are N-grams~\cite{santos_ngrams}. An N-gram is a contiguous sequence of n items from a given sequence of text. The rationale behind using N-grams to detect malware is that malicious software have structure, and N-grams work by capturing this structure (certain combinations of N-grams are more likely in samples belonging to some malware families than others. For the task of malware detection and classification, N-grams can be extracted from the hexadecimal representation of malware's binary content and from the assembly language source code, also known as byte N-grams and opcode N-grams, respectively. 

Byte N-grams and opcode N-grams refer to the unique combination of every n consecutive bytes and opcodes as individual features, respectively. As a result, N-gram based approaches construct a vector of features, where each feature in the vector indicates the number of appearances of a particular N-gram. Consequently, the length of the feature vector depends on the number of unique N-grams, which increases with N. This leads to two main shortcomings that limit the applicability of N-grams in a real-world scenario (when $N>3$). First, the resulting feature vector is very large as the model has to store the count for all N-grams occurring in the dataset. Second, the feature vector is very sparse, most features have zero values, increasing the space and time complexity of the resulting models. Generally, if there are too many features,the machine learning models tend to fit the noise in the training data. This is known as overfitting, and it results in poor generalization on newer data. As a result, machine learning approaches based on N-grams for the task of malware detection and classification have limited the size of N to size 3 or 4, and applied feature selection and dimensionality reduction techniques to reduce the dimensionality of the resulting feature vector.

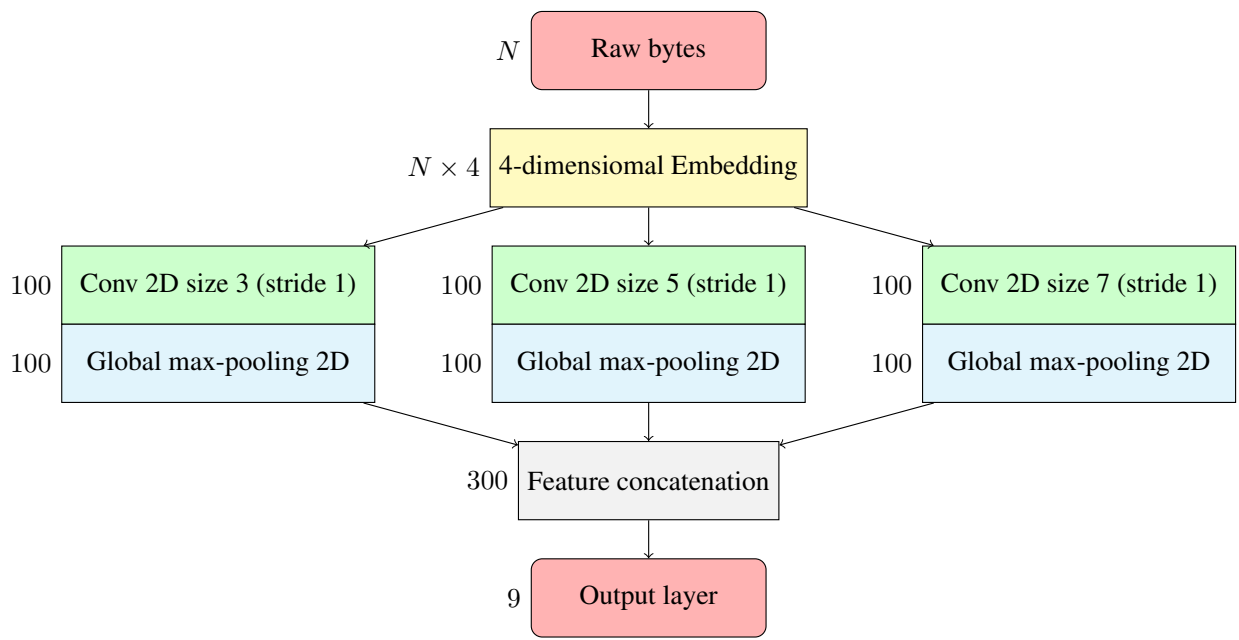
\begin{figure}[ht]
	\centering
	\tikzstyle{state}=[circle,
	thick,
	minimum size=1.2cm,
	draw=black,
	fill=white]
	\tikzstyle{input_output} = [rectangle, rounded corners, minimum width=3cm, minimum height=1cm,text centered, draw=black, fill=red!30]
	\tikzstyle{relu_act} = [rectangle, minimum width=1cm, minimum height=1cm,text centered, draw=black, fill=green!20]
	\tikzstyle{softmax} = [rectangle, minimum width=3cm, minimum height=1cm,text centered, draw=black, fill=red!30]
	\tikzstyle{selu} = [rectangle, minimum width=4cm, minimum height=1cm,text centered, draw=black, fill=blue]
	\tikzstyle{relu} = [rectangle, minimum width=4cm, minimum height=1cm,text centered, draw=black, fill=green!20]
	\tikzstyle{nonactivation} = [rectangle, minimum width=3cm, minimum height=1cm,text centered, draw=black, fill=gray!10]
	
	\tikzstyle{embedding} = [rectangle, minimum width=4cm, minimum height=1cm,text centered, draw=black, fill=yellow!30]
	\tikzstyle{pooling} = [rectangle, minimum width=4cm, minimum height=1cm,text centered, draw=black, fill=cyan!10]
	\resizebox{\columnwidth}{!}{%
		
		\begin{tikzpicture}[node distance=1.5cm]
		\node (raw_mnemonics) [input_output, label={left:$N$}] {Raw bytes};
		\node (emb) [embedding, below of=raw_mnemonics, label={left:$N \times 4$}] {4-dimensiomal Embedding};
		\node (conv_5) [relu, below of=emb, label={left:$100$}] {Conv 2D size 5 (stride 1)};
		\node (conv_3) [relu, left of=conv_5, label={left:$100$}, xshift=-4cm] {Conv 2D size 3 (stride 1)};
		\node (conv_7) [relu, right of=conv_5, label={left:$100$}, xshift=4cm] {Conv 2D size 7 (stride 1)};
		
		\node (pooling5) [pooling, below of=conv_5, label={left:$100$}, yshift=0.5cm] {Global max-pooling 2D};
		\node (pooling3) [pooling, below of=conv_3, label={left:$100$}, yshift=0.5cm] {Global max-pooling 2D};
		\node (pooling7) [pooling, below of=conv_7, label={left:$100$}, yshift=0.5cm] {Global max-pooling 2D};
		\node (concat) [nonactivation, below of=pooling5, label={left:$300$}] {Feature concatenation};
		\node (softmax) [input_output, below of=concat, label={left:$9$}] {Output layer};
		
		\draw[->] (raw_mnemonics) -- (emb);
		\draw[->] (emb) -- (conv_5);
		\draw[->] (emb) -- (conv_3);
		\draw[->] (emb) -- (conv_7);
		\draw[->] (pooling5) -- (concat);
		\draw[->] (pooling3) -- (concat);
		\draw[->] (pooling7) -- (concat);
		\draw[->] (concat) -- (softmax);
		
		\end{tikzpicture}
	}
	\caption{Shallow convolutional neural network architecture to extract N-gram like features from the hexadecimal representation of malware's binary content. Notice that the only difference with respect to the architecture that extracts N-gram like features from the assembly language source code of malware is the input. Instead of the raw bytes sequence, the input is the sequence of opcodes extracted from the assembly code.}
	\label{fig:shallow_cnn}
\end{figure}

Alternatively, Gibert et al.~\cite{21674592,GIBERT2021102159} proposed a shallow convolutional neural network architecture to extract N-gram like features from malware's assembly language source code and binary content. Cf. Figure~\ref{fig:shallow_cnn}. This is achieved by convolving various filters of different sizes k, where k $\in \{ 3, 5, 7 \}$ in our case, which indicates the number of opcodes and bytes to which is applied. Afterwards, a global max-pooling layer is applied to retrieve the maximum activation for each feature map independently of their position in the input sequence. This can be seen as if a particular N-gram has been found in the input sequence. Alternatively, one could employ a global average-pooling layer to retrieve the average of the activations for each feature map. However, in our experiments the global max-pooling layer achieved higher classification results. For more details about the architecture we refer the reader to the original publications~\cite{21674592,GIBERT2021102159}.

\subsection{Feature Fusion}
\begin{figure}[ht]
	\centering
	\resizebox{0.8\columnwidth}{!}{%
		
		\begin{tikzpicture}
		\usetikzlibrary{shapes.geometric, arrows.meta, backgrounds, fit, positioning}
		\usetikzlibrary{matrix,chains,scopes,positioning,arrows,fit}
		\tikzstyle{line} = [draw, -latex']
		\tikzstyle{feature_modality_1} = [circle, draw=black, fill=red]
		\tikzstyle{feature_modality_2} = [circle, draw=black, fill=green]
		\tikzstyle{feature_modality_3} = [circle, draw=black, fill=blue!70]
		\tikzstyle{black_dot} = [circle, draw=black, fill=black,inner sep=1pt]
		
		\tikzstyle{feature_vector_1} = [rectangle, rounded corners, minimum width=2.5cm, minimum height=1cm, text centered, draw=black, fill=red!30]
		
		\tikzstyle{feature_vector_2} = [rectangle, rounded corners, minimum width=2.5cm, minimum height=1cm, text centered, draw=black, fill=green!30]
		
		\tikzstyle{feature_vector_3} = [rectangle, rounded corners, minimum width=2.5cm, minimum height=1cm, text centered, draw=black, fill=blue!30]
		
		\tikzstyle{feature_fusion} = [rectangle, rounded corners, minimum width=8cm, minimum height=1cm, text centered, draw=black, fill=orange!40]

		\tikzstyle{classifier} = [rectangle, text centered, draw=black, fill=gray!20, minimum width=2.5cm, minimum height=1cm]
		
		\node (vector1) [feature_vector_1,label=above:Feature vector $\overrightarrow{v}$] {};
		\node (f1) [feature_modality_1, xshift=-1cm] {};
		\node (f2) [feature_modality_1, right of=f1, xshift=-0.5cm] {};
		\node (f3) [feature_modality_1, right of=f2, xshift=-0.5cm] {};
		\node (dot1) [black_dot, right of=f3, xshift=-0.5cm, yshift=-0.1cm] {};
		\node (dot2) [black_dot, right of=dot1, xshift=-0.8cm] {};
		\node (dot3) [black_dot, right of=dot2, xshift=-0.8cm] {};
		
		\node (vector2) [feature_vector_2,label=above:Feature vector $\overrightarrow{w}$, right of=vector1, xshift=2.5cm] {};
		\node (f4) [feature_modality_2, left of=vector2] {};
		\node (f5) [feature_modality_2, right of=f4, xshift=-0.5cm] {};
		\node (f6) [feature_modality_2, right of=f5, xshift=-0.5cm] {};
		\node (dot4) [black_dot, right of=f6, xshift=-0.5cm, yshift=-0.1cm] {};
		\node (dot5) [black_dot, right of=dot4, xshift=-0.8cm] {};
		\node (dot6) [black_dot, right of=dot5, xshift=-0.8cm] {};
		
		\node (doti1) [black_dot, right of=vector2, xshift=0.75cm,] {};
		\node (doti2) [black_dot, right of=doti1, xshift=-0.8cm] {};
		\node (doti3) [black_dot, right of=doti2, xshift=-0.8cm] {};
		
		\node (vector3) [feature_vector_3,label=above:Feature vector $\overrightarrow{z}$, right of=doti3, xshift=0.75cm] {};
		\node (f7) [feature_modality_3, left of=vector3] {};
		\node (f8) [feature_modality_3, right of=f7, xshift=-0.5cm] {};
		\node (f9) [feature_modality_3, right of=f8, xshift=-0.5cm] {};
		\node (dot7) [black_dot, right of=f9, xshift=-0.5cm, yshift=-0.1cm] {};
		\node (dot8) [black_dot, right of=dot7, xshift=-0.8cm] {};
		\node (dot9) [black_dot, right of=dot8, xshift=-0.8cm] {};
		
		\node (early_fusion) [feature_fusion, below of=vector2, yshift=-1cm] {};
		
		\node (f11) [feature_modality_1, xshift=-1.5cm, below of=vector2, yshift=-1cm, xshift=-2cm] {};
		\node (f21) [feature_modality_1, right of=f11, xshift=-0.5cm] {};
		\node (f31) [feature_modality_1, right of=f21, xshift=-0.5cm] {};
		\node (dot11) [black_dot, right of=f31, xshift=-0.5cm, yshift=-0.1cm] {};
		\node (dot21) [black_dot, right of=dot11, xshift=-0.8cm] {};
		\node (dot31) [black_dot, right of=dot21, xshift=-0.8cm] {};
		
		\node (f41) [feature_modality_2, right of=dot31, yshift=0.1cm, xshift=-0.3cm] {};
		\node (f51) [feature_modality_2, right of=f41, xshift=-0.5cm] {};
		\node (f61) [feature_modality_2, right of=f51, xshift=-0.5cm] {};
		\node (dot41) [black_dot, right of=f61, xshift=-0.5cm, yshift=-0.1cm] {};
		\node (dot51) [black_dot, right of=dot41, xshift=-0.8cm] {};
		\node (dot61) [black_dot, right of=dot51, xshift=-0.8cm] {};
		
		\node (f71) [feature_modality_3, right of=dot61, yshift=0.1cm, xshift=-0.3cm] {};
		\node (f81) [feature_modality_3, right of=f71, xshift=-0.5cm] {};
		\node (f91) [feature_modality_3, right of=f81, xshift=-0.5cm] {};
		\node (dot71) [black_dot, right of=f91, xshift=-0.5cm, yshift=-0.1cm] {};
		\node (dot81) [black_dot, right of=dot71, xshift=-0.8cm] {};
		\node (dot91) [black_dot, right of=dot81, xshift=-0.8cm] {};
		
		\node (classifier) [classifier, below of=early_fusion, yshift=-1cm] {Classifier};
		
		\node (benign) [feature_modality_2, below of=classifier, yshift=-0.5cm, xshift=-0.3cm,label=left:benign] {};
		\node (malicious) [feature_modality_1, below of=classifier, yshift=-0.5cm, xshift=0.3cm,label=right:malware] {};
		
		\draw[->] (f1) -- (f11);
		\draw[->] (f2) -- (f21);
		\draw[->] (f3) -- (f31);
		
		\draw[->] (f4) -- (f41);
		\draw[->] (f5) -- (f51);
		\draw[->] (f6) -- (f61);
		
		\draw[->] (f7) -- (f71);
		\draw[->] (f8) -- (f81);
		\draw[->] (f9) -- (f91);
		
		\draw[->] (early_fusion) -- (classifier);
		\draw[->] (classifier) -- (benign);
		\draw[->] (classifier) -- (malicious);
		
		\end{tikzpicture}
	}
	\caption{Early fusion strategy}
	\label{fig:early_fusion}
\end{figure}
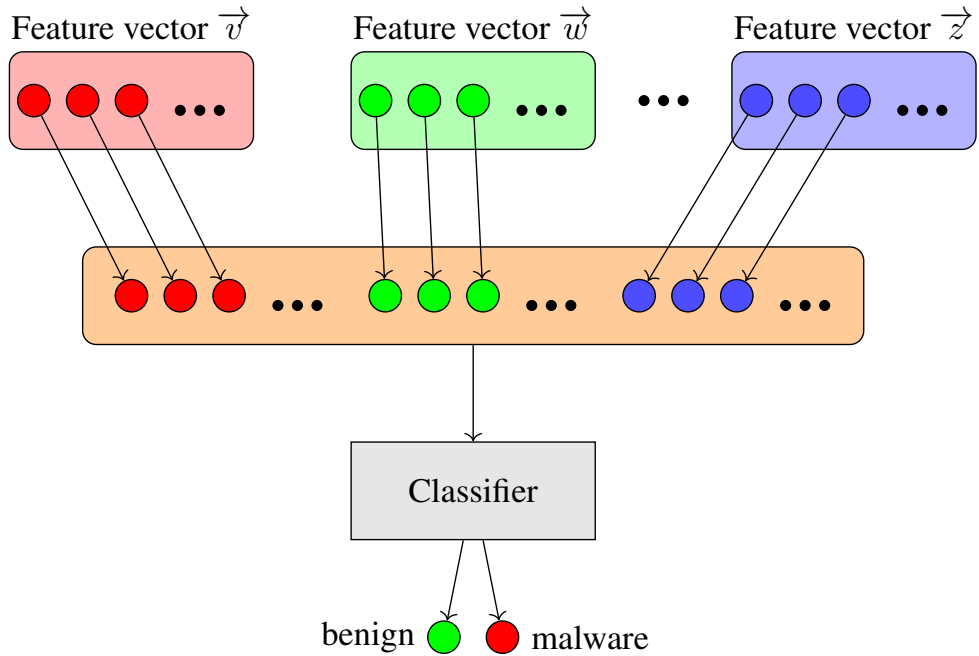

At this point, the executable is represented as various feature vectors $\overrightarrow{v}$, $\overrightarrow{w}$, $\overrightarrow{x}$, ..., $\overrightarrow{z}$, one for each type of features, e.g. BYTE\_MD, BYTE\_1G, BYTE\_ENT, etc, that provide an abstract view of their content.
To integrate the feature vectors, our system employs an early fusion mechanism to create a joint representation of the features from multiple modalities. This fusion mechanism combines the various feature vectors by concatenating them into a single feature vector. Cf. Figure~\ref{fig:early_fusion}. Afterwards, a single model is trained to learn the correlation and interactions between the features of each modality. Given various feature vectors $\overrightarrow{v}$, $\overrightarrow{w}$, $\overrightarrow{x}$, ..., $\overrightarrow{z}$ containing the different types of features, the prediction of the final model, denoted as $h$, can be written as:
$$ p = h\left ( [\overrightarrow{v}, \overrightarrow{w}, ..., \overrightarrow{z}] \right ) = \left ( [v_{1}, v_{2}, ..., v_{i}, w_{1}, w_{2}, ..., w_{j}, ..., z_{1}, z_{2}, ..., z_{k}] \right )$$

\subsection{Gradient Boosting Trees}
\label{sec:gbt}
For classification purposes, we trained our machine learning model using XGBoost~\cite{DBLP:journals/corr/ChenG16}, a parallel implementation of the gradient boosting tree classifier. The impact of XGBoost has been widely recognized in many machine learning and data mining competitions, where approximately half of the winning solutions used XGBoost.
Next, we review gradient boosting tree algorithms. For a more detailed description of gradient boosting trees and XGBoost we refer the reader to the original article~\cite{DBLP:journals/corr/ChenG16}.

Boosting is an ensemble learning technique for building a strong classifier in the form of an ensemble of weak classifiers. When the weak learners are decision trees, the resulting algorithm is called boosted trees. Gradient Boosting Trees (GBT) is an ensemble learning technique that builds one tree at a time, where each new tree attempts to minimize the errors of the previous tree. This is achieved by combining sequentially the weak learners in a way that each new learner fits to the residuals from the previous steps so that the model improves. Afterwards, the final model aggregates the results from each step to build the strong learner. Notice that to detect the residuals a loss function should be used, e.g. the logarithmic loss for classification tasks. Subsequently, adding many trees sequentially, and each one focusing in the errors from the previous one, makes boosting a highly efficient and accurate model for classification tasks.

\section{Evaluation} 
\label{sec:evaluation}
\subsection{The Microsoft Malware Classification Challenge Dataset}
The task of Windows malware detection has not received the same attention by the research community as other domains, where rich datasets exist~\cite{GIBERT2020102526}. In addition, the copyright laws that prevent sharing benign software has exacerbated this situation. As a result, no dataset containing benign and malicious software is available to the public for research. The only benchmark available for malware detection is the Ember dataset~\cite{DBLP:journals/corr/abs-1804-04637}, which provides a collection of features from PE files. However, the raw binaries are not included and thus, the application of deep learning or the extraction of new features from the executables is not possible. To make things worse, even as malicious binaries may be obtained for internal use through web services such as VirusTotal, its subsequent sharing of the binary or the vendor antimalware labels assigned are prohibited. Furthermore, unlike other domains where data samples may be labeled quickly and, in most cases by a non-expert, determining if a file is malicious or not is a very time consuming and complex process, even for security experts. This issues makes it impossible to meaningfully compare accuracy across works because different datasets with distinct labeling procedures are used from one work to another. For this reason, instead of creating our own private dataset, we decided to evaluate our system with the data provided by Microsoft for the Big Data Innovators Gathering Challenge of 2015, the only high-quality public labeled benchmark available for malware classification research~\cite{DBLP:journals/corr/abs-1802-10135}. 

This dataset has become the standard benchmark to evaluate machine learning approaches for malware classification and currently, it is publicly available in the Kaggle platform~\footnote{\url{https://www.kaggle.com/c/malware-classification}}. It includes half a terabyte of data consisting of 10868 samples for training and 10873 samples for testing. To sum up, the dataset contains samples representing 9 different malware families, where each sample has associated two files: (1) the hexadecimal representation of malware's binary content and (2) its corresponding assembly language source code, generated using the IDA Pro disassembler~\footnote{\url{https://www.hex-rays.com/products/ida/}}. Cf. Figures~\ref{fig:hex_view} and~\ref{fig:assembly_view}. The families represented are the following: (1) Ramnit, (2) Lollipop, (3) Kelihos\_ver3, (4) Vundo, (5) Simda, (6) Tracur, (7) Kelihos\_ver1, (8) Obfuscator.acy and (9) Gatak.
Following, a brief description of the behavior of each family is provided according to Microsoft Security Intelligence~\footnote{\url{https://www.microsoft.com/en-us/wdsi}}.
\begin{description}
	\item[Ramnit.] This worm family is reported to have the capability to steal your sensitive information such as saved FTP credentials and browser cookies, and can spread via removable drives.
	\item[Lollipop.] This adware family is reported to display ads in your browser as you navigate the Internet. In addition, it can also redirect your search engine results, monitor your PC and download other applications.
	\item[Kelihos\_ver3.] This backdoor family is reported to install on the system in order to download other components, and includes a backdoor that gives the attacker further control over the affected system. Computer systems affected by this malware were used as bots in the Kelihos botnet, now deceased. Version 3 of the software.
	\item[Vundo.] This trojan family is reported to cause popups and advertising for rogue anti-spyware programs.
	\item[Simda.] This backdoor family is reported to install on the system to give an attacker remote control of the system. In addition, this malware is reported to steal personal and system data, take screenshots and download additional files.
	\item[Tracur.] This family of malware is reported to redirect you Internet search queries to malicious URLs to download and install other malware.
	\item[Kelihos\_ver1.] This backdoor family is reported to install on the system in order to download other components, and includes a backdoor that gives the attacker further control over the affected system. Computer systems affected by this malware were used as bots in the Kelihos botnet, now deceased. Version 1 of the malicious software.
	\item[Obfuscator.ACY.] This family of malware comprises software that has been obfuscated, that is, software that has tried to hide its behavior or purpose so that anti-malware engines do not detect it. The software that lies underneath this obfuscation can have any purpose.
	\item[Gatak.] This trojan family is reported to silently download and install other software without the user's consent.
	
\end{description}

As it can be observed in Table~\ref{tab:microsoft_training_distribution}, the dataset is very imbalanced and the class distribution is not uniform among families, i.e. the number of samples belonging to some families significantly outnumber the number of samples belonging to the remaining families.

\begin{table}[h]
	\centering
	\caption{Class distribution in the Microsoft dataset~\cite{DBLP:journals/corr/abs-1802-10135}}
	\label{tab:microsoft_training_distribution}
	\begin{tabular}{l|r|l}
		\hline
		Family         & \#Samples   & Type               \\ \hline
		Ramnit         & 1541        & Worm               \\
		Lollipop       & 2478        & Adware             \\
		Kelihos\_ver3  & 2942        & Backdoor           \\
		Vundo          & 475         & Trojan             \\
		Simda          & 42          & Backdoor           \\
		Tracur         & 751         & TrojanDownloader   \\
		Kelihos\_ver1  & 398         & Backdoor           \\
		Obfuscator.ACY & 1228        & Obfuscated malware \\
		Gatak          & 1013        & Trojan           \\ \hline
	\end{tabular}
\end{table}
\subsection{Experimental Setup}
The system has been deployed on a machine with an Intel Core i7-7700k CPU, 2xGeforce GTX1080Ti GPUs and 64Gb RAM. To implement the convolutional neural network architectures it has been used Tensorflow~\cite{tensorflow2015-whitepaper}.

The generalization performance of our multimodal approach has been estimated using k-fold cross validation, with k equals to 10. K-fold cross validation is a model validation technique to assess how accurately a predictive model will perform in practice. In k-fold validation, the dataset is partitioned into k folds of equal size. Then, the following procedure is followed for each one of the k folds:
\begin{itemize}
	\item A model is trained using $k-1$ of the folds as training data.
	\item The resulting model is validated on the remaining fold of the data.
\end{itemize}
Afterwards, the performance measure reported is the average of the values achieved for each fold. 

Next, are presented the performance metrics and the experiments carried out to evaluate our multimodal approach. In particular, the experiments have been designed to evaluate the classification performance of each individual subset of features, to compare the performance of the model with and without hand-crafted and deep features, and to evaluate the fusion of features using early fusion and the forward stepwise selection algorithm. Lastly, our approach has been compared with the state-of-the-art approaches in the literature.

\subsubsection{Performance Metrics}
Regarding the performance metrics used to evaluate our approach, we will report two metrics, the accuracy and the logarithmic loss.

The accuracy is simply the fraction of correct predictions. Formally, accuracy is defined as follows:
$$ accuracy = \frac{Number\; of\; correct\; predictions}{Total\; number\; of\; predictions}$$
However, accuracy alone is not a good evaluation metric to assess the robustness of machine learning models in datasets where there exist a large class imbalance. Subsequently, the multi-class logarithmic loss (logloss) has been used to to assess the performance of the predictions. The logarithmic loss is the cross entropy between the distribution of true labels and the predicted probabilities. Formally, it is defined as follows:
$$logloss = -\frac{1}{N} \sum_{i=1}^{N} \sum_{j=1}^{M} y_{i,j}\;log(p_{i,j})$$
where $N$ is the number of observations, $M$ is the number of class labels, log is the natural logarithm, $y_{i,j}$ is 1 if the observation $i$ is in class $j$ and 0 otherwise, and $p_{i,j}$ is the predicted probability that observation i is in class j. Notice that for the test set it is only provided the multi-class logarithmic loss or logloss. This is because the labels of the samples in the test set are not provided and to assess the performance of a given model you need to submit the predicted probabilities to Kaggle.

\subsubsection{Individual Subsets of Features Performance Analysis}
Table~\ref{tab:individual_feature_subsets} presents the 10-fold cross validation accuracy and logloss achieved by the machine learning models trained using only an individual subset of features in the training and test sets. XGboost has various hyperparameters that are completely tunable. Cf. Table~\ref{tab:hyperparameters}. For now, we will skip the details of the hyperparameters and set all of them to their baseline value. Table~\ref{tab:individual_feature_subsets} gives us some clues about the discriminative power of each feature category. There are three feature categories that have very low accuracy and high logarithmic loss in comparison to the remaining ones. The feature categories are the following: \emph{BYTE\_MD}, \emph{ASM\_MD}, and \emph{ASM\_PIXEL}. This means that this type of features alone are not enough to correctly categorize malware into families with high accuracy, but as described in Section~\ref{sec:results_early_fusion}, some of them are very valuable in the construction of the multimodal boosted decision trees. Thus, we decided to keep this features in our final model. However, the classification performance of the models trained on a single feature category cannot compete with the state-of-the-art approaches in the literature. Thus, in the following sections various mechanisms to fuse the different feature categories are studied. See Section~\ref{sec:state_of_the_art}. 

\begin{table}[h]
	\centering
	\caption{Classification performance of each feature category}
	\label{tab:individual_feature_subsets}
		\begin{tabular}{l|cc|l}
			\hline
			& \multicolumn{2}{c|}{Train} & Test    \\ \hline
			Feature Category  & Accuracy     & Logloss     & Logloss \\ \hline
			BYTE\_MD          & 0.8634   & 0.4066   & 0.3975 \\
			BYTE\_1G          & 0.9835   & 0.0602   & 0.0386 \\
			BYTE\_ENT         & 0.9778   & 0.0788   & 0.0728 \\
			BYTE\_HARALICK    & 0.9741   & 0.0892   & 0.0833 \\
			BYTE\_LBP         & 0.9791   & 0.0782   & 0.0608 \\
			BYTE\_IMG\_CNN    & 0.9464   & 0.0139   & 0.1066 \\
			BYTE\_ENT\_CNN    & 0.9703   & 0.1118   & 0.1291 \\
			BYTE\_NGRAMS\_CNN & 0.9756   & 0.0074   & 0.0302 \\
			ASM\_MD           & 0.9072   & 0.3323   & 0.2807 \\
			ASM\_OPC          & 0.9598   & 0.0090   & 0.0263 \\
			ASM\_PIXEL        & 0.7661   & 0.5346   & 0.5442 \\
			ASM\_REG          & 0.9334   & 0.0168   & 0.0533 \\
			ASM\_SYM          & 0.8905   & 0.0288   & 0.0942 \\
			ASM\_API          & 0.9840   & 0.0591   & 0.0546 \\
			ASM\_DD           & 0.9851   & 0.0606   & 0.0411 \\
			ASM\_SEC          & 0.9879   & 0.0749   & 0.0329 \\
			ASM\_MISC         & 0.9926   & 0.0282   & 0.0206 \\
			ASM\_NGRAMS\_CNN  & 0.9917   & 0.0299   & 0.0356 \\ \hline    
		\end{tabular}
\end{table}

One benefit of gradient boosting trees is that in order to construct the decision trees, they have to explicitly calculate the importance for each feature in the dataset. Feature importance is a score that indicates how useful or valuable each feature has been in the construction of the boosted decision trees. The more a particular feature has been selected to make key decisions in the decision trees, the higher its relative importance. For a single decision tree, importance is calculated by the amount each feature split point improves the performance measure, weighted by the number of observations the node is responsible for. In the case of XGBoost, this performance measure is the purity (Gini index) used to select the split points. These feature importances are then averaged across all of the decision trees within the model to retrieve their final importance scores. Accordingly, we included the top 20 most important features for each feature category in~\ref{appendix:feature_importance_by_category}.

\begin{table*}[h]
	\centering
	\caption{List of hyperparameters and their values}
	\label{tab:hyperparameters}
	\resizebox{\textwidth}{!}{%
		\begin{tabular}{l|l|l|l}
			\hline
			Hyperparameter     & Baseline Value & Best Value & Description                                                                             \\ \hline
			eta                & 0.2            & 0.1          & Step size shrinkage.                                                                    \\
			max\_depth         & 5              & 3          & Maximum depth of a tree.                                                                \\
			gamma              & 0.0            & 0.0        & Minimum loss reduction required to make a further partition on a leaf node of the tree. \\
			min\_child\_weight & 1              & 1          & Minimum sum of instance weight (hessian) needed in a child.                             \\
			colsample\_bytree  & 1.0            & 1.0        & It is the subsample ratio of columns when constructing each tree.                       \\
			subsample          & 1.0            & 1.0        & Subsample ratio of the training instances.                                              \\ \hline
		\end{tabular}%
	}
\end{table*}

\subsubsection{Early-fusion Performance}
\label{sec:results_early_fusion}

This second experiment aims at studying the performance of the boosted trees models when trained with various feature categories combined using early-fusion with and without the deep features. Subsequently, we trained the following models:
\begin{description}
	\item[Hex-based hand-crafted features.] It refers to the model trained using the features manually extracted from the hexadecimal view of malware. 
	Feature categories: \emph{BYTE\_MD}, \emph{BYTE\_1G}, \emph{BYTE\_ENT}, \emph{BYTE\_HARALICK}, \emph{BYTE\_LBP}.
	\item[Hex-based hand-crafted \& deep features.] It refers to the model trained using both the hand-crafted and deep features extracted from the hexadecimal view of malware. Feature categories: \emph{BYTE\_MD}, \emph{BYTE\_1G}, \emph{BYTE\_ENT}, \emph{BYTE\_LBP}, \emph{BYTE\_HARALICK}, \emph{BYTE\_IMG\_CNN}, \emph{BYTE\_ENT\_CNN}, \emph{BYTE\_NGRAMS\_CNN}.
	\item[Assembly-based hand-crafted features.] It refers to the model trained using the features manually extracted from the assembly view of malware. Feature categories: \emph{ASM\_MD}, \emph{ASM\_OPC}, \emph{ASM\_PIXEL}, \emph{ASM\_REG}, \emph{ASM\_SYM}, \emph{ASM\_API}, \emph{ASM\_DD}, \emph{ASM\_SEC}, \emph{ASM\_MISC}.
	\item[Assemly-based hand-crafted \& deep features.] It refers to the model trained using both the hand-crafted and deep features extracted from the assembly view of malware. Feature categories: \emph{ASM\_MD}, \emph{ASM\_OPC}, \emph{ASM\_PIXEL},\\* 
	\emph{ASM\_REG}, \emph{ASM\_SYM}, \emph{ASM\_API}, \emph{ASM\_DD}, \emph{ASM\_SEC}, \emph{ASM\_MISC}, \emph{ASM\_NGRAMS\_CNN}.
	\item[Hand-crafted features.] It refers to the model trained using the features manually extracted from both the hexadecimal and assembly view of malware. Feature categories: \\* 
	\emph{BYTE\_MD}, \emph{BYTE\_1G}, \emph{BYTE\_ENT}, \emph{BYTE\_HARALICK},\\* \emph{BYTE\_LBP}, \emph{ASM\_MD}, \emph{ASM\_OPC}, \emph{ASM\_PIXEL}, \emph{ASM\_REG}, \emph{ASM\_SYM}, \emph{ASM\_API}, \emph{ASM\_DD}, \emph{ASM\_SEC}, \emph{ASM\_MISC}.  
	\item [Deep features.] It refers to the model trained using the features transferred from the deep learning models. Feature categories:  \emph{BYTE\_IMG\_CNN}, \emph{BYTE\_ENT\_CNN},\\* \emph{BYTE\_NGRAMS\_CNN}, \emph{ASM\_NGRAMS\_CNN}.
	\item [Hand-crafted \& deep features.] It refers to the model trained using both the hand-crafted and deep features extracted from the hexadecimal and the assembly view of malware. Feature categories: \emph{BYTE\_MD}, \emph{BYTE\_1G}, \emph{BYTE\_ENT}, \emph{BYTE\_HARALICK}, \emph{BYTE\_LBP}, \emph{ASM\_MD}, \emph{ASM\_OPC},\\* \emph{ASM\_PIXEL}, \emph{ASM\_REG}, \emph{ASM\_SYM}, \emph{ASM\_API}, \emph{ASM\_DD}, \emph{ASM\_SEC}, \emph{ASM\_MISC}, \emph{BYTE\_IMG\_CNN}, \emph{BYTE\_ENT\_CNN}, \emph{BYTE\_NGRAMS\_CNN}, \emph{ASM\_NGRAMS\_CNN}.
\end{description}
\begin{table}[h]
	\centering
	\caption{Classification performance of the early-fusion models.}
	\label{tab:early_fusion_results}
		\begin{tabular}{l|cc|l}
			\hline
			& \multicolumn{2}{c|}{Train} & Test    \\ \hline
			Feature Category                             & Accuracy     & Logloss     & Logloss \\ \hline
			Hex-based hand-crafted features              & 0.9962       & 0.0138   & 0.0207         \\
			Hex-based hand-crafted \& deep features      & 0.9958       & 0.0154   & 0.0193         \\
			Assembly-based hand-crafted features         & 0.9978       & 0.0100   & 0.0063         \\
			Assembly-based hand-crafted \& deep features & 0.9978       & 0.0069   & 0.0052 \\
			Hand-crafted features                        & 0.9976       & 0.0088   & 0.0073 \\
			Deep features                                & 0.9954       & 0.0154   & 0.0194 \\
			Hand-crafted \& deep features                & 0.9987 & 0.0059 & 0.0086 \\\hline
		\end{tabular}
\end{table}

Table~\ref{tab:early_fusion_results} presents the classification performance of the aforementioned methods. It can be observed that the models trained with both hand-crafted and deep features achieve higher accuracy and lower logarithmic loss than their hand-crafted counterpart. However, the more complex model, the lower its performance. It can be observed that the model trained with hand-crafted and deep features from both the hexadecimal and the assembly view of malware, contains 956 more features than the hand-crafted model, making it more prone to overfitting. Overfitting occurs when the classification model models too well the training data, that is, the details and noise in the training data, to the extent that it negatively impacts the performance of the model on new data (fails to generalize to unseen data). 
For this reason, the hyperparameters of the XGBoost models were tuned to avoid overfitting. In particular, we performed an heuristic search over the hyperparameters listed in Table~\ref{tab:hyperparameters}. Following, is provided a brief description of each hyperparameter:
\begin{itemize}
	\item eta. Eta is the learning rate of our gradient boosted trees model. It indicates how much we update the prediction with each successive tree. The lower the eta is, the more conservative the boosting process will be.
	\item max\_depth. It refers to the maximum depth of a tree.The large max\_depth is, the more complex and likely to overfit the model will be.
	\item gamma. The gamma is the minimum loss reduction required to make a further partition on a leaf node of the tree. A larger gamma makes the algorithm more conservative.
	\item min\_child\_weight. It refers to the minimum sum of instance weight needed in a leaf. If the weight for every instance is 1, it directly indicates to the minimum number of instances needed in a node.
	\item colsample\_bytree. It indicates the subsample ratio of features when constructing each tree.
	\item subsample. It refers to the subsample ratio of the training instances. For instance, setting its value to 0.5 means that XGBoost would randomly sample half of the training data prior to growing trees.
\end{itemize}
\begin{figure*}
	\centering
	\begin{subfigure}{.48\textwidth}
		\centering
		\includegraphics[width=.9\columnwidth]{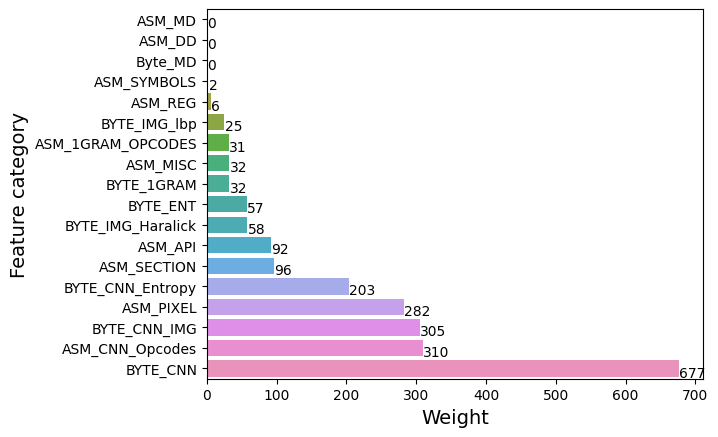}
		\caption{Number of times a feature category has been used to split the data across all trees.}
		\label{fig:feature_category_importance_weight}
	\end{subfigure}%
	\hfill
	\begin{subfigure}{.48\textwidth}
		\centering
		\includegraphics[width=.9\columnwidth]{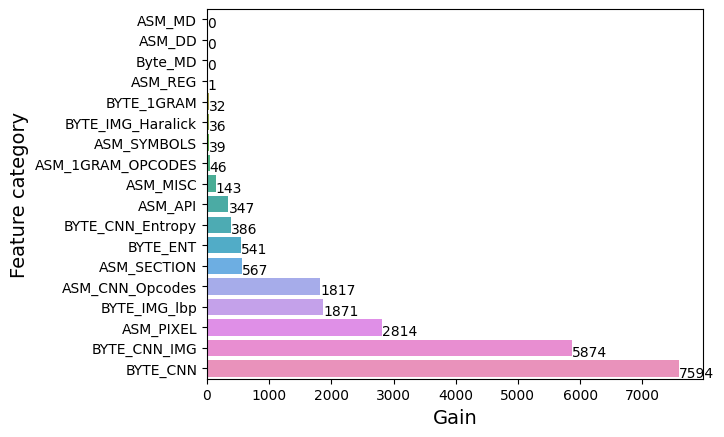}
		\caption{Average gain across all splits the feature categories have been used in.}
		\label{fig:feature_category_importance_gain}
	\end{subfigure}
	\caption{Importance of each feature category.}
	\label{fig:feature_category_importance}
\end{figure*}
By training the boosted trees model using the best hyperparameters defined in Table~\ref{tab:hyperparameters} we have been able to fine-tune the algorithm and reduce the logarithmic loss achieved on the test set from 0.008860 to 0.00636, the second lowest loss reported in any published article by a single model so far, only behind the model trained using both the hand-crafted and deep features extracted from the assembly view of malware, which achieved a logloss of 0.00516. Notice that state-of-the-art approaches in the literature~\cite{10.1145/2857705.2857713,7847046} employ one or more ensemble learning techniques such as bagging. However, the fact that the model trained with only the assembly features achieves better results than the model using all features indicates that there are at least one or more subsets of features that do not apply to new data and negatively affect the model's ability to generalize. In addition, as it can be observed in Figure~\ref{fig:feature_category_importance}, the number of times a particular feature occurs in the trees of the model greatly varies between categories. For instance, the model trained with all features has used features from the \emph{BYTE\_CNN}, \emph{BYTE\_CNN\_IMG}, \emph{ASM\_CNN\_OPCODES} and \emph{ASM\_PIXEL} categories 840, 291, 310 and 300 times, respectively. On the other hand, metadata features (\emph{ASM\_MD} and \emph{BYTE\_MD}) and  data define features (\emph{ASM\_DD}) have not been used for building the final model. However, it is not possible to know if it is because of their unimportance or because the combination of other features achieves the same effect.
Given this circumstances, we decided to use various feature selection techniques to select the most discriminant subset of features.

\subsubsection{Feature Selection}
Fusing all subsets of features into a single feature vector produces a feature vector of size equals to 3349, quite large compared to the number of training instances, which is 10868. However, as it can be observed in Section~\ref{sec:results_early_fusion}, a feature vector with all features is not optimal and produces an overfitted model, i.e. a model with poor generalization performance. What is happening is known as the Hughes phenomenon. That is, as the number of features grows, the classifier's performance increases until it reaches the optimal number of features. Afterwards, the performance of the classifier decreases as the dimensionality of the feature space increases, under the assumption that the number of training samples remains fixed. 

In theory, the classification algorithm used to train our models (see Section~\ref{sec:gbt}) naturally selects which features are most important when constructing the decision trees. However, as it can be observed in Section~\ref{sec:results_early_fusion}, when the model is trained using all features it ends up overfitting the training data and performing poorly on the test set. In consequence, univariate feature selection has been investigated to select a subset of the features that have a major statistically significant relationship with the target variable (the malware families) based on various univariate statistical tests. More specifically, the best subset of features has been selected using the Chi-squared~\cite{doi:10.1080/14786440009463897}, the ANOVA F-value~\cite{STHLE1989259}, and the Mutual Information~\cite{10.1093/bioinformatics/18.S231} score functions. Unfortunately, as it can be observed in Table~\ref{tab:K_features_results}, the models trained using univariate feature selection performed poorly in comparison to the models presented in Section~\ref{sec:results_early_fusion}. Our intuition is that, as univariate feature selection works by selecting the features that have a significant statistical relationship with the target variable (class or family) and does not take into account the relationship between features, it might be discarding features that alone are not very discriminant but work well in combination with other features. In consequence, a wrapper method to select the best subset of features has been proposed.

\begin{table*}[ht]
	\centering
	\caption{List of models trained using the K best features according to various univariate metrics and their evaluation with XGBoost.}
	\label{tab:K_features_results}
	\begin{tabular}{l|cc|l|cc|l|cc|l}
		\hline
		& \multicolumn{3}{c|}{Chi square}       & \multicolumn{3}{c|}{ANOVA f-value}     & \multicolumn{3}{c}{Mutual Information} \\ \hline
		& \multicolumn{2}{c|}{Train} & Test     & \multicolumn{2}{c|}{Train} & Test      & \multicolumn{2}{c|}{Train} & Test       \\ \hline
		K    & Accuracy     & Logloss     & Logloss  & Accuracy     & Logloss     & Logloss   & Accuracy     & Logloss     & Logloss    \\ \hline
		20   & 0.4987       & 1.3673      & 1.3654   & 0.6138       & 0.9078      & 0.9094    & 0.9961       & 0.0163      & 0.0124    \\
		50   & 0.5627       & 1.2434      & 1.2304   & 0.6138       & 0.9086      & 0.9092    & 0.9969       & 0.0135      & 0.0100    \\
		100  & 0.9544       & 0.1323      & 0.1355   & 0.6137       & 0.9087      & 0.9089    & 0.9967       & 0.0126      & 0.0089    \\
		200  & 0.9977       & 0.0095      & 0.0097   & 0.7187       & 0.6430      & 0.6430    & 0.9970       & 0.0116      & 0.0092    \\
		500  & 0.9980       & 0.0071      & 0.0097   & 0.9956       & 0.0144      & 0.0196    & 0.9975       & 0.0108      & 0.0096    \\
		1000 & 0.9985       & 0.0065      & 0.0117   & 0.9983       & 0.0068      & 0.0096    & 0.9983       & 0.0073      & 0.0079    \\
		1500 & 0.9984       & 0.0061      & 0.0094   & 0.9985       & 0.0061      & 0.0089    & 0.9983       & 0.0059      & 0.0088    \\
		2000 & 0.9988       & 0.0062      & 0.0093   & 0.9986       & 0.0057      & 0.0088    & 0.9983       & 0.0056      & 0.0084    \\
		2500 & 0.9988       & 0.0057      & 0.0085   & 0.9988       & 0.0058      & 0.0097    & 0.9987       & 0.0058      & 0.0073    \\ \hline
	\end{tabular}
\end{table*}

\subsubsection{Forward Stepwise Selection Technique}
Starting with a model containing no features, the forward stepwise selection algorithm gradually increases the feature set by adding subsets of features, one by one. To determine the subset of features to add at each step, it is used the logarithmic loss achieved on the validation set using K-fold cross validation, where K equals 3. That is, at each step, we add to the feature set the subset of features that produce the minimum value of the logarithmic loss. This process stops when adding more subsets of features does not decrease the value of the logarithmic loss. The rationale behind using a smaller K than in Section~\ref{sec:results_early_fusion} is that it greatly reduces the training computational time as it reduces to three the number of models to be trained. 
\begin{table*}[h]
	\centering
	\caption{Gradual addition of feature categories using the forward stepwise selection algorithm.}
	\label{tab:forward_stepwise_results}
	\begin{tabular}{l|c|cc|l}
		\hline
		\multicolumn{2}{c|}{} & \multicolumn{2}{c|}{Train} & Test    \\ \hline
		Feature Category & Number of features & Accuracy     & Logloss     & Logloss \\ \hline
		C1: ASM\_NGRAMS\_CNN       & 300   & 0.9928 & 0.0270 & 0.0356 \\
		C2: C1 + ASM\_PIXEL        & 1100  & 0.9979 & 0.0079 & 0.0072 \\
		C3: C2 + BYTE\_IMG\_CNN    & 1356  & 0.9983 & 0.0071 & 0.0110 \\
		C4: C3 + ASM\_API          & 2150  & 0.9986 & 0.0065 & 0.0094\\
		C5: C4 + BYTE\_1G          & 2405  & 0.9989 & 0.0061 & 0.0101\\ 
		C6: C5 + BYTE\_NGRAMS\_CNN & 2705  & 0.9989 & 0.0060 & 0.0092 \\ 
		C7: C6 + ASM\_OPC          & 2798  & 0.9987 & 0.0059 & 0.0092 \\ 
		C8: C7 + BYTE\_LBP         & 2906  & 0.9987 & 0.0058 & 0.0091 \\ 
		C9: C8 + ASM\_MD           & 2908  & 0.9987 & 0.0057 & 0.0076 \\ 
		C10: C9 + ASM\_REG         & 2946  & 0.9988 & 0.0057 & 0.0081\\ \hline
	\end{tabular}
\end{table*}
Unfortunately, as it can be observed in Table~\ref{tab:forward_stepwise_results}, although adding more subsets of features decrease the logarithmic loss on the validation set, when the model is evaluated on the test set, better results have been achieved by a model with a smaller subset of features. In this case, the lowest logarithmic loss on the test set has been achieved by a model trained using only two feature subsets: (1) \emph{ASM\_NGRAMS\_CNN}, and (2) \emph{ASM\_PIXEL}. 


\subsubsection{Comparison with the State-Of-The-Art}
\label{sec:state_of_the_art}
Given that previous feature selection techniques have not achieved better results than the model trained using only the assembly features, we built the final model using as features those extracted from the assembly language source code, \emph{ASM\_MD}, \emph{ASM\_OPC}, \emph{ASM\_PIXEL}, \emph{ASM\_REG}, \emph{ASM\_SYM}, \emph{ASM\_API},\\* \emph{ASM\_DD}, \emph{ASM\_SEC}, \emph{ASM\_MISC}, \emph{ASM\_NGRAMS\_CNN}, plus three feature subsets from the binary content, \emph{BYTE\_MD},\\* \emph{BYTE\_1G}, and \emph{BYTE\_NGRAMS\_CNN}. The reason is that adding any other of the feature subsets deteriorates the classification performance of the system.

Next, our approach is compared with the state-of-the-art methods in the literature that are based on feature engineering and deep learning. Only approaches that have evaluated their methods using the Microsoft benchmark and employ K-fold cross validation have been selected for comparison. 
This has been done to ensure fairness when comparing the methods. Take into account that methods evaluated on different datasets are not directly comparable, and most often than not, the source code is not available online or it does not work properly, i.e. the parameters have been optimized for their own dataset, there are missing libraries, etc. For this reason, the results presented in Table~\ref{tab:state_of_the_art_results} are those published in their original publications, without modifications. 

\begin{table*}[h]
	\centering
	\caption{Comparison with the state-of-the-art methods on the Microsoft Malware Classification Challenge benchmark. Those approaches that their authors have not tested their performance on the test set or did not make public the K-fold cross validation accuracy or logarithmic loss appear with a ‘-’ mark. Approaches with a ‘*’ mark indicate that they performed 5-fold cross validation instead of 10-fold cross validation. Approaches that have not performed K-fold cross validation but have used a single hold-out validation set are marked with "**". }
		\label{tab:state_of_the_art_results}
		\begin{tabular}{l|l|cc|l}
			\hline
			& & \multicolumn{2}{c|}{10-fold Cross Validation} & Test    \\ \hline
			Method & Input  & Accuracy     & Logloss     & Logloss \\ \hline
			Narayanan et al.~\cite{7856826} & Grayscale images  & 0.9660  & -  & - \\
			Kebede et al.~\cite{8268747}** & Grayscale images & 0.9915 & - & - \\
			Gibert et al.~\cite{Gibert2018} & Grayscale images  &  0.9750    &   -   &  0.1845 \\
			Kalash et al.~\cite{8328749}** & Grayscale images & 0.9852 & - & 0.0571 \\
		    Liu	et al.~\cite{8610223} & Grayscale images        & 0.9900  & -  & - \\
			Vinayakumar et al.~\cite{8681127} & Grayscale images & 	0.9630 & - & - \\
			Lo et al.~\cite{8763852} & Grayscale images & - & -& 0.0361 \\
			Qiao et al.~\cite{8887383} & Grayscale images & 0.9876 & - & - \\
			Sudhakar and Kumar\cite{SUDHAKAR2021334} & Grayscale images & 0.9853 & - & - \\
			Xiao et al.~\cite{XIAO2021102420} & Grayscale images & 0.9894 &  - & - \\
			Çayır et al.~\cite{CAYIR2021102133}** & Grayscale images & 0.9926 & - & - \\
			Lin and Yeh\cite{math10040608} & Grayscale images & 0.9632 & - & - \\
			Yuan et al.~\cite{YUAN2020101740} & Markov images & 0.9926 & 0.0518 & - \\
			Kim et al.~\cite{KIM201883} & RGB images  & 0.9574  & - & - \\
			Jiang et al.~\cite{10.1007/978-3-030-36711-4_14} & RGB images & 0.9973 & - & 0.0220 \\
			Zhang et al.~\cite{pr9060929} & RGB images & 0.9934 & - & - \\
			Gibert et al.~\cite{AAAI1816133} & Structural entropy   &  0.9828    &  -    &  0.1244\\
			Xiao et al.~\cite{XIAO202049} & Structural entropy        &  0.9972 & - & 0.0314 \\
			Yan et al.~\cite{8809504} & Control flow graph & 0.9925 & 0.0543 & - \\
			Hu et al.~\cite{10.1147/JRD.2016.2559378} & Opcode 4-grams     & 0.9930 & - & 0.0546\\
			Gibert et al.~\cite{21674592} & Opcode sequence      & 0.9917 & - & 0.0244 \\
			Gibert et al.~\cite{DBLP:conf/ijcnn/GibertMP19} & Opcode sequence  &  0.9913 &    -   &  0.0419 \\
			McLaughlin et al.~\cite{10.1145/3029806.3029823} & Opcode sequence      & 0.9903   &  -   & 0.0515 \\
			Yousefi-Azar et al.~\cite{7966342} & Byte sequence      &  0.9309   &  - & -\\
			Drew et al.~\cite{10.1186/s13635-017-0055-6} & Byte sequence &  0.9741 & - & 0.0479\\
			Raff et al.~\cite{DBLP:conf/aaai/RaffBSBCN18} & Byte sequence      & 0.9641  &  - & 0.3071  \\
			Krčál et al.~\cite{krcal2018deep} & Byte sequence      &  0.9756   & -   & 0.1602  \\
			Kim and Cho\cite{KIM2022102501} & Byte sequence & 0.9747 & - & - \\
			Le et al.~\cite{LE2018S118} & Compressed byte sequence      &  0.9820   &  -  & 0.0774 \\
			Gibert et al.~\cite{DBLP:conf/icann/GibertMP18} & Compressed byte sequence      &   0.9861   &     -        &  0.1063 \\
			Messay-Kebede et al.~\cite{8556722} & Opcode statistics and byte sequence & 0.9907 & - & - \\
			Gibert et al.~\cite{9206671} & Opcode and byte sequences      &   0.9924          &    -         &   -      \\
			Gibert et al.~\cite{GIBERT2020101873} & API calls, Opcode and byte sequences      & 0.9975    &    -         &   -      \\
			Gao et al.~\cite{GAO2020102661}* & Hand-crafted features & 0.9969 & - & - \\
			Ahmadi et al.~\cite{10.1145/2857705.2857713}* & Hand-crafted features      &   0.9977          &   0.0096           &   0.0063      \\
			Zhang et al.~\cite{7847046} & Hand-crafted features    &  0.9976           &      -       &  0.0042      \\ \hline
			\textbf{Proposed system} &  \textbf{Hand-crafted and deep features}           & \textbf{0.9981}    & \textbf{0.0070} &  \textbf{0.0040}  \\ \hline
		\end{tabular}
	
\end{table*}

As it can be observed, our multimodal approach achieves the highest accuracy and lowest logarithmic loss in the training set using K-fold cross validation, and the lowest logarithmic loss in the test set, outperforming any machine learning approach presented in the literature so far. This has been possible thanks to the N-gram like features extracted by the shallow convolutional neural networks from both the hexadecimal representation of malware's binary content and its assembly language source code. This features are not only discriminant but also computationally inexpensive in comparison to traditional N-gram features, which require to manually enumerate all N-grams during training, and later reduce and select a smaller subset of characterizing N-grams by employing feature selection or dimensionality reduction techniques. In addition, our approach outperformed state-of-the-art approaches without employing sampling techniques, stacking and ensemble learning techniques as in Zhang et al.~\cite{10.1145/2857705.2857713} and Ahmadi et al.~\cite{7847046}.

\section{Conclusions}
\label{sec:conclusions}
In this paper, we present a novel multimodal approach that combines feature engineering and deep learning to extract features from both the hexadecimal representation of malware's binary content and its assembly language source code to achieve state-of-the-art performance in the task of malware classification. To the best of our knowledge, this research is the first application to combine feature engineering and deep learning using a simple, but yet effective, early fusion mechanism for the problem of malware classification. The success of our multimodal approach would not had been possible without the extraction of the N-gram like features using the shallow convolutional neural networks trained on the bytes and opcodes sequence representing malware's binary content and its assembly language source code, respectively, which provide a computational inexpensive way to extract long N-gram like features without having to exhaustively enumerate all N-grams during training. Furthermore, by fusing hand-crafted features and deep features we have been able to combine the strengths of both approaches, descriptive domain-specific features and the ability of deep learning to automatically learn to extract features from sequential data without relying on the experts' knowledge of the domain.

Reported results allow to assess its effectiveness with respect to the state-of-the-art for the task of malware classification, achieving the highest accuracy and lowest logarithmic loss reported in the Microsoft benchmark so far, while only using a single model to generate the final predictions. 

\subsection{Future Work}
One future line of research could be the design and development of new architectures to extract features from malware's representation as grayscale images or its structural entropy as they perform below average in comparison with other features. In addition, a second line of research could be the analysis of the performance of various texture pattern extractors, to complement the Haralick and Local Binary Pattern feature extractors. Lastly, a third line of research could be the study of ensemble learning techniques such as blending and bagging, to build a stronger classifier by combining multiple models. In addition, the following limitations and open research questions should be dealt with in order to address the threat of malware.

	\subsection{Limitations of Deep Learning}
	With the vertical (numbers and volumes) and horizontal ( types and functionality) expansion of malware threats during the last decade, malware detection has remained a hot topic as the number of yearly publications addressing the task demonstrate and it is still far from being solved. Recently, deep learning has started being adopted because of its ability to extract features from raw data~\cite{DBLP:conf/aaai/RaffBSBCN18,krcal2018deep}. However, its application to the task of malware detection and classification has been limited so far due to the computational resources required to train such models. Take into account that depending on the type of input we might end up dealing with sequences of millions of time steps, which far exceeds the length of the input of any previous deep learning sequence classifier, leading to a huge consumption of the GPU memory in the first convolutional layers and the use of large filter widths and strides to balance the memory usage and the computational workload. This has forced researchers to explore alternatives to reduce and compress the information in the bytes sequence to make it a manageable problem~\cite{DBLP:journals/virology/BaysaLS13,AAAI1816133,10.1145/2016904.2016908,LE2018S118}.
	
	In addition, contrarily to computer vision, where the features extracted by deep learning aim to replace previous feature extractors, for the problem of malware detection the features learned through deep learning greatly differ from those usually extracted by domain experts to provide an abstract representation of the executable. More specifically, deep learning is only capable to substitute N-gram features, one of the many types of features usually extracted to characterize the executables. In consequence, machine learning systems trained on hand-engineered features tend to outperform deep learning approaches. For an extended description of common features extracted by domain experts we refer the readers to the work of Anderson et al.~\cite{DBLP:journals/corr/abs-1804-04637} and the references therein.
	
	\subsection{Open Research Questions}
	Persistent research efforts have been made using machine learning to address the threat of malware. However, machine learning models have proved to be susceptible to adversarial examples and concept drift. 
	
	On the one hand, adversarial examples are inputs to a machine learning model that have been specifically crafted to exploit the weakness of the ML model in order to incorrectly label the malicious sample as benign. For instance, Hu et al.~\cite{DBLP:journals/corr/HuT17} proposed to craft the adversarial examples using a generative adversarial network (GAN) architecture named MalGAN, which learns which API imports should be added to the original sample to bypass detection. Sucio et al.~\cite{suciu2019exploring} proposed to modify the existing bytes of a binary by appending content of benign executables and by using the Fast Gradient Method~\cite{DBLP:journals/corr/GoodfellowSS14} to modify the random bytes appended at the end of the executables until the adversarial example evades detection. Demetrio et al.~\cite{9437194} proposed a functionality-preserving black-box optimization attack based on the injection of content of benign executables. To do so, they either inject code at the end of the file or within some newly-created section.
	Demetrio et al.~\cite{10.1145/3473039} proposed three functionality-preserving attacks against ML-based models. These attacks, named Full DOS, Extend, and Shift, inject a crafted payload by manipulating the DOS header, extending it, and shifting the content in the first section, respectively. Given the success of the aforementioned attacks, countermeasures must be developed in other to mitigate the threat of malware. One mitigation strategy may be to consider features that are not affected by the changes at the section or byte level. Another countermeasure may be adversarial retraining. That is, including adversarial examples generated by the aforementioned attacks on the training set. 
	
	On the other hand, concept drift refers to the problem of changing underlying relationships in the data over time. Traditional machine learning methods assume that the mapping learned from historical data will be valid for new data in the future and the relationships between input and output do not change over time. However, malware is pushed to evolve in order to evade detection, making this assumption false. Moreover, malicious software, as any other piece of software, naturally evolve over time due to changes resulting from adding features, fixing bugs, porting to new environments, etcetera. Therefore, the performance of any ML-based detector naturally degrades over time as malware evolves. To tackle concept drift, Jordaney et al.~\cite{203684} proposed a framework to identify aging classification models during deployment. This is done by assessing the quality of the prediction made by the machine learning model. It builds and make use of p-values to compute a per-class threshold to identify unreliable predictions. The idea behind is to detect when the model's performance starts to degrade before it actually happens. Contrarily, Pendlebury et al.~\cite{235493} proposed a set of space and time constraints that eliminate "spatial bias", training and testing distributions not representative of the real-world data, and "temporal bias", incorrect time splits of training and testing sets.
	In addition, they introduced a set of time-aware performance metrics that allows for the comparison of different classifiers while considering time decay. Following the aforementioned research, new approaches might be developed to deal with concept drift by classification with rejection, in which examples that have drifted far away from the training distribution, and thus are likely to be misclassified by the ML models, are quarantined.

\section*{Acknowledgements}
This project has received funding from Enterprise Ireland, the Spanish Science and Innovation Ministry funded project PID2019-111544GB-C22, the European Union’s Horizon 2020 Research and Innovation Programme under the Marie Skłodowska-Curie grant agreement No 847402. The views and conclusions contained in this document are those of the authors and should not be interpreted as representing the official policies, either expressed or implied, of CeADAR, University College Dublin, and the University of Lleida.

\section*{Conflicts of Interest}
The authors declare that they have no known competing financial interests or personal relationships that may appear to influence the work reported in this paper.

\bibliographystyle{unsrt}  
\bibliography{refs}

\appendix
\section{Feature Importance by Category}
\label{appendix:feature_importance_by_category}
Following, the readers will find the top 20 features of each feature category. The importance of each feature is defined as the average gain across all splits the feature has been used in.
\begin{figure}[H]
	\centering
	\includegraphics[width=\columnwidth]{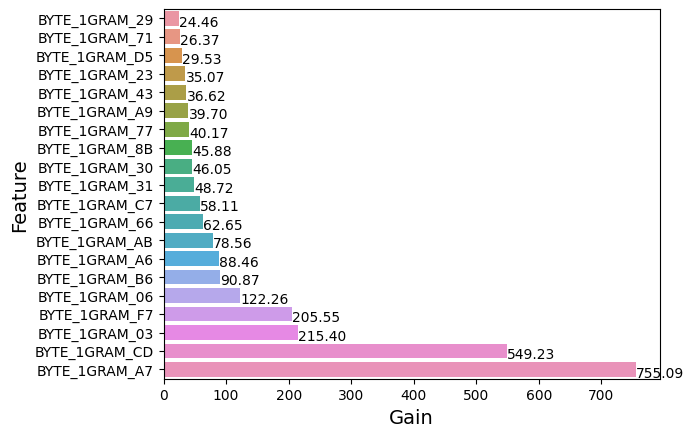}
	\caption{Top 20 \emph{BYTE\_1G} features.}
	\label{fig:feature_category_importance_byte_1g}
\end{figure}

\begin{figure}[H]
	\centering
	\includegraphics[width=\columnwidth]{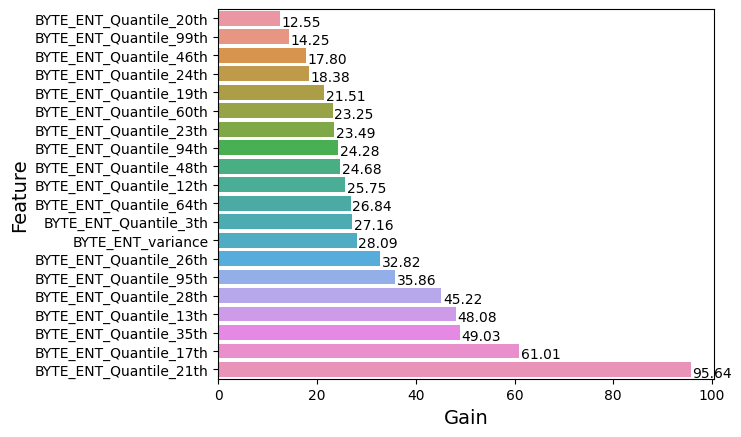}
	\caption{Top 20 \emph{BYTE\_ENT} features.}
	\label{fig:feature_category_importance_byte_ent}
\end{figure}
\begin{figure}[H]
	\centering
	\includegraphics[width=\columnwidth]{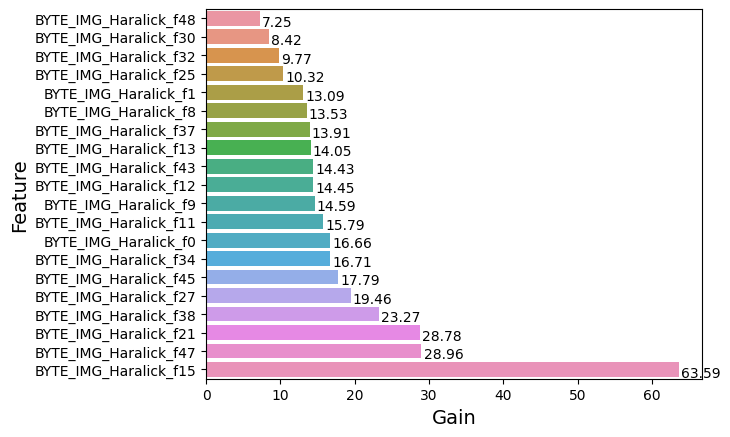}
	\caption{Top 20 \emph{BYTE\_IMG\_HAR} features.}
	\label{fig:feature_category_importance_byte_img_har}
\end{figure}
\begin{figure}[H]
	\centering
	\includegraphics[width=\columnwidth]{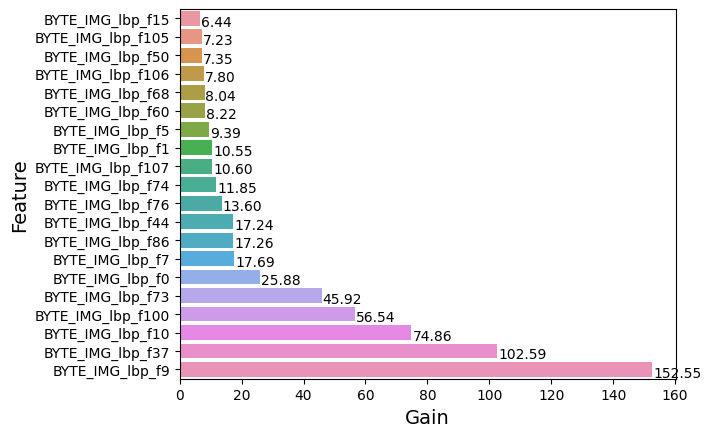}
	\caption{Top 20 \emph{BYTE\_IMG\_LBP} features.}
	\label{fig:feature_category_importance_byte_img_lbp}
\end{figure}
\begin{figure}[H]
	\centering
	\includegraphics[width=\columnwidth]{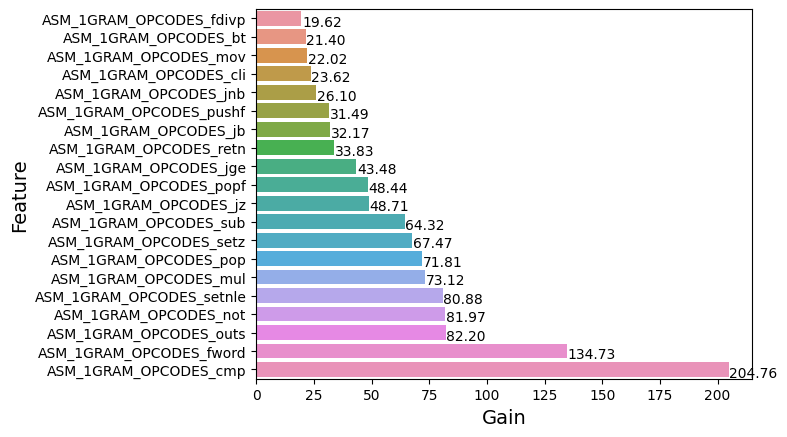}
	\caption{Top 20 \emph{ASM\_OPC} features.}
	\label{fig:feature_category_importance_asm_opc}
\end{figure}

\begin{figure}[H]
	\centering
	\includegraphics[width=\columnwidth]{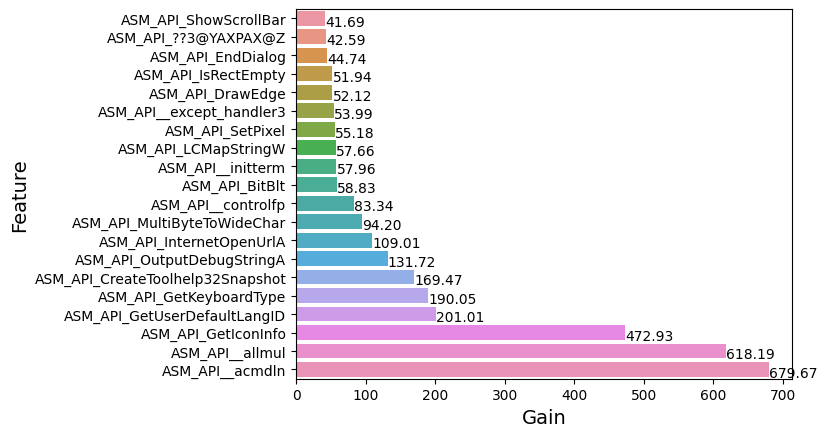}
	\caption{Top 20 \emph{ASM\_API} features.}
	\label{fig:feature_category_importance_asm_api}
\end{figure}

\begin{figure}[H]
	\centering
	\includegraphics[width=\columnwidth]{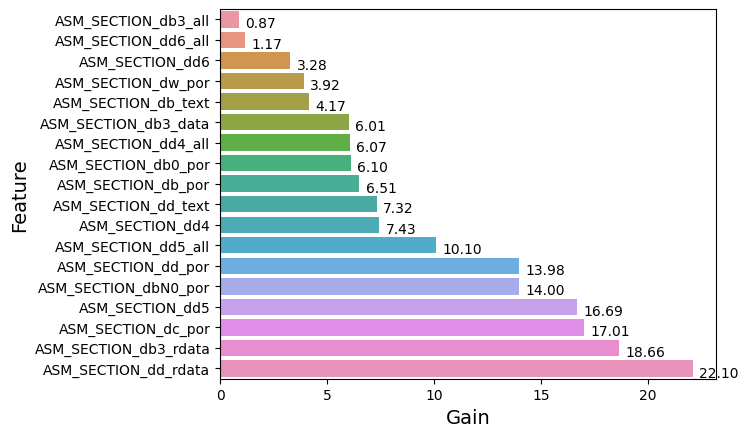}
	\caption{Top 20 \emph{ASM\_DD} features.}
	\label{fig:feature_category_importance_asm_dd}
\end{figure}

\begin{figure}[H]
	\centering
	\includegraphics[width=\columnwidth]{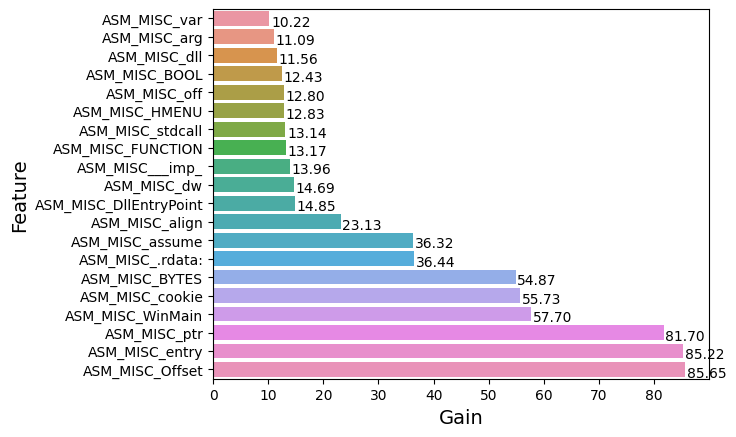}
	\caption{Top 20 \emph{ASM\_MISC} features.}
	\label{fig:feature_category_importance_asm_misc}
\end{figure}

\begin{figure}[H]
	\centering
	\includegraphics[width=\columnwidth]{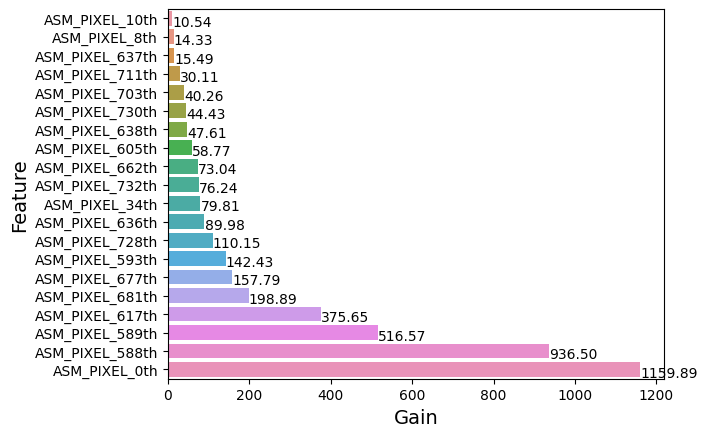}
	\caption{Top 20 \emph{ASM\_PIXEL} features.}
	\label{fig:feature_category_importance_asm_pixel}
\end{figure}

\begin{figure}[H]
	\centering
	\includegraphics[width=\columnwidth]{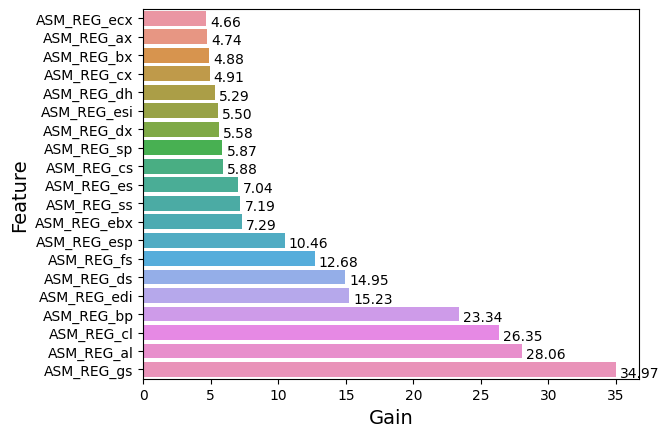}
	\caption{Top 20 \emph{ASM\_REG} features.}
	\label{fig:feature_category_importance_asm_reg}
\end{figure}

\begin{figure}[H]
	\centering
	\includegraphics[width=\columnwidth]{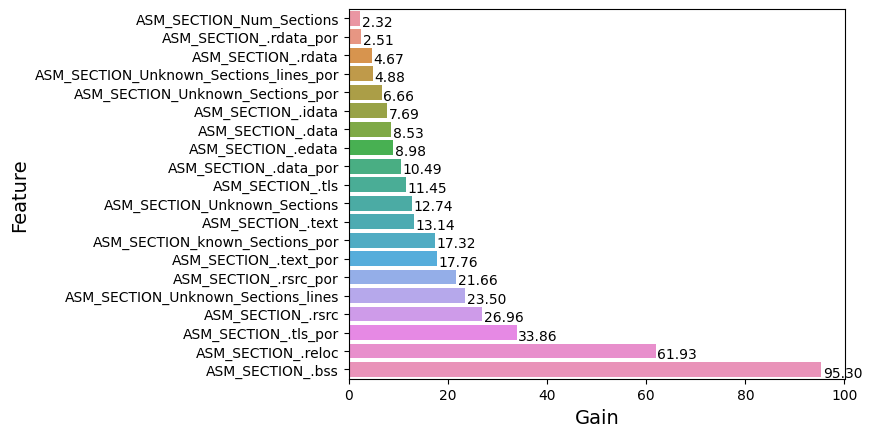}
	\caption{Top 20 \emph{ASM\_SEC} features.}
	\label{fig:feature_category_importance_asm_sec}
\end{figure}

\begin{figure}[H]
	\centering
	\includegraphics[width=\columnwidth]{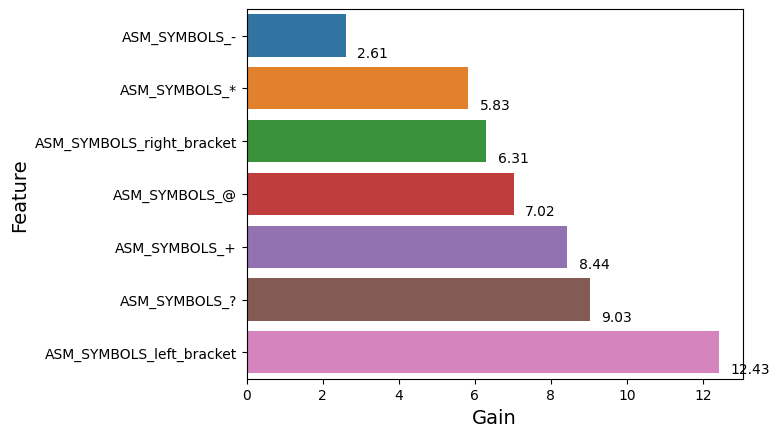}
	\caption{\emph{ASM\_SYM} feature importance.}
	\label{fig:feature_category_importance_asm_sym}
\end{figure}

\end{document}